\documentclass[fleqn,usenatbib]{mnras}
\usepackage{newtxtext,newtxmath}

\usepackage[T1]{fontenc}
\usepackage{ae,aecompl}
\usepackage{color}

\usepackage{graphicx}	
\usepackage{amsmath}	
\usepackage{amssymb}	

\title[Non-spherical ULA dark matter haloes in the dSphs]{Non-sphericity of ultralight-axion dark matter haloes in the Galactic dwarf spheroidal galaxies}

\author[K. Hayashi \& I. Obata]{
Kohei Hayashi$^{1}$\thanks{E-mail: hayaipmu@icrr.u-tokyo.ac.jp}
and Ippei Obata$^{1}$\thanks{E-mail: obata@icrr.u-tokyo.ac.jp}
\\
$^{1}$Institute for Cosmic Ray Research, The University of Tokyo}

\date{Accepted 2019 October~11. Received 2019 October 10; in original form 2019 February 8}

\pubyear{2019}

\begin{document}
\label{firstpage}
\pagerange{\pageref{firstpage}--\pageref{lastpage}}
\maketitle

\begin{abstract}
Ultralight-axion (ULA) dark matter is one of the possible solutions to resolve small-scale problems, especially the core-cusp problem.
This is because ULA dark matter can create a central soliton core in all dark matter haloes stemmed from the quantum pressure against gravity below the de Broglie wavelength, which becomes manifest on astrophysical scales with axion mass range $\sim10^{-22}$~eV.
In this work, we apply our non-spherical dynamical models to the kinematic data of eight classical dwarf spheroidals (dSphs) to obtain more reliable and realistic limits on ULA particle mass.
This is motivated by the reasons that the light distributions of the dSphs is not spherical, nor are the shapes of dark matter haloes predicted by ULA dark matter simulations. 
Compared with the previous studies on ULA dark matter assuming spherical mass models, our result is less stringent than those constraints due to the uncertainties on non-sphericity.
On the other hand, remarkably, we find that the dSphs would prefer to have a flattened dark matter halo rather than a spherical one, especially Draco favours a strongly elongated dark matter halo caused naively by the assumption of a soliton-core profile.
Moreover, our consequent non-spherical core profiles are much more flattened than numerical predictions based on ULA dark matter, even though there are still uncertainties on the estimation of dark matter halo structure.
To alleviate this discrepancy, further understanding of baryonic and/or ULA dark matter physics on small mass scales might be needed.
\end{abstract}



\section{Introduction}

In recent decades, precision observation cosmology has revealed that dark matter constitutes $\sim26$ per cent of the total energy density in the Universe~\citep{2016A&A...594A..13P,2018arXiv180706209P}. The concordant $\Lambda$ cold dark matter~(CDM) theory gives an excellent description of the cosmological and astrophysical observations on large spatial scales such as cosmic microwave background~(CMB) and large-scale structure of galaxies~\citep[e.g.,][]{2006Natur.440.1137S,2006PhRvD..74l3507T,2013ApJS..208...20B}. Meanwhile, on galactic and sub-galactic scales there have been several long-standing tensions between the predictions from pure dark matter simulations based on this model and some observational facts~\citep[see][]{2015PNAS..11212249W,2017ARA&A..55..343B}. In particular, there are major three challenges in $\Lambda$CDM. (i) ``missing satellite problem'': $N$-body simulations based on $\Lambda$CDM predict that Milky Way~(MW) sized dark haloes have significantly more subhaloes than the observed number of dwarf satellites~\citep{1999ApJ...524L..19M, 1999ApJ...522...82K}. Recently, owing to the wide and deep imaging surveys, a large number of new ultra-faint dwarf galaxies are discovered~\citep[e.g.,][]{2006ApJ...647L.111B,2007ApJ...654..897B,2015ApJ...807...50B,2015ApJ...805..130K,2016ApJ...832...21H,2018PASJ...70S..18H}. 
In total, $\sim50$ satellites were excavated so far, but this problem still remains.
(ii) ``core-cusp problem'': dark matter haloes from $\Lambda$CDM simulations are predicted to  have strongly cusped central density profiles~\citep[e.g.,][]{1997ApJ...490..493N,1997ApJ...477L...9F,2013ApJ...767..146I}, while observations of low mass galaxies are suggested to have cored dark halo profiles~\citep[e.g.,][]{2001MNRAS.323..285B,2007ApJ...663..948G,2008AJ....136.2761O,2010AdAst2010E...5D}. (iii) ``Too-big-to-fail problem'': the most massive $\Lambda$CDM subhaloes associated with a MW-sized dark matter halo are much more concentrated than the dark matter haloes in the dwarf galaxies within the MW~\citep{2011MNRAS.415L..40B,2012MNRAS.422.1203B}, M31~\citep{2014MNRAS.440.3511T}, and the Local Group~\citep{2014MNRAS.444..222G}.

One of the possible mechanisms to resolve the above problems is to consider baryonic physics such as the stellar feedback from supernova explosions and stellar winds~\citep[e.g.,][]{1996MNRAS.283L..72N,2002MNRAS.333..299G,2012MNRAS.422.1231G,2014MNRAS.437..415D}, the ram pressure stripping, the tidal stripping~\citep[e.g.,][]{2012ApJ...761...71Z,2014ApJ...786...87B,2016MNRAS.457.1931S}, the interaction between dark matter and baryons through dynamical friction~\citep[e.g.,][]{ElZant:2001re,Ma:2004nw,Cole:2011yp,Inoue:2011nn,Nipoti:2014xha,2014JCAP...04..021D,2014JCAP...12..051D,2018PhRvD..98f3517D} and the photoionization of neutral gas with UV background~\citep[e.g.,][]{1977MNRAS.179..541R,1992MNRAS.256P..43E,2000ApJ...539..517B}. These mechanisms can have an impact on dark halo structures and suppress the galaxy formation in dark haloes.
Another solution is, more radically, to develop alternative dark matter scenarios, such as warm dark matter~\citep[e.g.,][]{2001ApJ...556...93B,2005PhRvD..71f3534V,2012MNRAS.420.2318L,2017PhRvD..96b3522I}, self-interacting dark matter~\citep[e.g.,][]{1992ApJ...398...43C,2000PhRvL..84.3760S,2016PhRvL.116d1302K,2018PhR...730....1T}, strong interacting massive particle~\citep[e.g.,][]{2014arXiv1402.5143H,2015PhRvL.115b1301H}, and axion like particles~\citep[e.g.,][]{2014NatPh..10..496S,2016PhR...643....1M}.
These dark matter models also suppress the matter power spectrum on small scales and create a cored density profile without relying on any baryonic physics, because dark matter particles based on these models retain higher velocity than that on CDM models and thus smear structures on small scales.~\citep[see][for a comprehensive review]{2003Sci...300.1909O}.
In this paper, we focus on the latter approach, especially an extremely light scalar bosonic particle.

It is known that string theory predicts a plentitude of axion-like particles with very light mass range~\citep{Svrcek:2006yi, Arvanitaki:2009fg,2018EPJWC.16806005L} and it behaves like a non-relativistic fluid in our Universe.
Hereafter we call them ``ultralight axion" (ULA)\footnote{ULA has been used in previous researches as one of the conventional phrases in the nomenclature of ultralight scalar dark matter with mass $m \sim 10^{-22}\text{eV}$ as well as "Fuzzy dark matter", "Wave dark matter", and so on. Although there is actually no explicit distinction between them in our work, we have adopted this phrase from the point of axion cosmology.}.
Its wave properties potentially resolve the small scale problems of CDM on sub-galactic scales~\citep{2000PhRvL..85.1158H, 2006PhLB..642..192A, 2010PhRvD..82j3528M}.
ULA dark matter predicts that dark matter distributions on large spatial scales are resemblant to those from $\Lambda$CDM models~\citep{1993ApJ...416L..71W, 2014PhRvD..90b3517U, 2014NatPh..10..496S}. On the other hand, its de Broglie wavelength with the ULA mass $m_\psi \sim 10^{-22}\text{eV}$ becomes relevant on astrophysical scales, hence consequently quantum pressures suppress the formation of low-mass haloes below $\sim10^{10}M_{\odot}$ and predict the kpc-scale central density cores in dSph-sized subhaloes~\citep{2000PhRvL..85.1158H,2014MNRAS.437.2652M,2014NatPh..10..496S,2016PhRvD..94d3513S, 2016PhRvD..94l3523V,2017MNRAS.471.4559M,2017PhRvD..95d3541H}.
For the above reasons, ULA dark matter has become one of the promising candidates for dark matter.
 
 The particle mass of ULA dark matter, $m_\psi$, is one of important parameters to help set a core radius and central density of ULA dark matter halo structures~\citep{2014NatPh..10..496S}. Furthermore, constraining on its particle mass range plays an indispensable role in probing ULA dark matter through experiments \citep[e.g.][]{Zioutas:2004hi, Vogel:2013bta, Bahre:2013ywa, 2015PhRvL.115t1301S,Kahn:2016aff, 2017arXiv170806367A,Obata:2018vvr} or astrophysical phenomena \citep[e.g.][]{Angus:2013sua, Khmelnitsky:2013lxt, Payez:2014xsa, Conlon:2015uwa, Aoki:2016kwl, Berg:2016ese, Marsh:2017yvc,2018arXiv181103525F,2019PhRvD..99f3517V}.
 Over a dozen works have made an attempt to set the constraints on the ULA particle mass by using various independent observations such as the substructures in the MW~\citep[e.g.,][]{2012JCAP...02..011L,2017MNRAS.468.1338C,2018arXiv180800464A}, the rotation curves of low surface brightness galaxies~\citep[e.g.,][]{2018MNRAS.475.1447B,2019MNRAS.488.5127F}, and the CMB measurements~\citep[e.g.,][]{2015MNRAS.450..209B,2018MNRAS.476.3063H}~\footnote{For further discussions about the constraints on the ULA dark matter particle mass, we will describe in Section~4.1.}.

The dwarf spheroidal~(dSph) galaxies in the MW are excellent laboratories to shed light on the fundamental properties of dark matter because these galaxies are the most dark matter dominated system with dynamical mass-to-light ratios of 10 to 1000~\citep{2012AJ....144....4M}.
The dSph galaxies are in close to the Sun, so that their line-of-sight velocities for their resolved member stars can be measured by stellar spectroscopy~\citep[e.g.,][]{2007ApJS..171..389W}.
Therefore, this kinematic information offers us an opportunity to scrutinize dark halo structures in the dSph galaxies. 
Actually, it has been revealed that some of the dSphs have cored dark matter density profiles as a result of the application of dynamical modelings to the kinematic data of the dSphs~\citep{2008ApJ...681L..13B,2011ApJ...742...20W,2012MNRAS.419..184A}, although it is still an ongoing debate~\citep{2010MNRAS.408.2364S,2014arXiv1406.6079S,2018MNRAS.481..860R}.
Moreover, the kinematic data of the classical dSphs enables us to determine ULA particle mass~$m_\psi$ based on the Jeans analysis with assuming a ULA dark halo model predicted by numerical simulations~\citep{2014NatPh..10..496S,2014PhRvL.113z1302S,2015MNRAS.451.2479M,2017MNRAS.468.1338C,2016MNRAS.460.4397C,2017MNRAS.472.1346G}

All these studies have assumed, however, that the luminous and dark components are spherically symmetric, despite the facts that the stellar distributions of the dSphs are actually non-spherical~\citep{2012AJ....144....4M} and some dark matter simulations predict non-spherical virialized dark haloes~\citep{2002ApJ...574..538J,2007ApJ...671.1135K,2009ApJ...697..850W,2014MNRAS.439.2863V}.
Motivated by this reason, \citet{2012ApJ...755..145H} constructed the non-spherical dynamical mass models for the dSphs based on the axisymmetric Jeans equations. They applied these models to the line-of-sight velocity dispersion profiles of six bright dSphs in the MW to constrain their non-spherical dark halo structures and concluded that most of the dSphs suggest to have very flattened dark haloes~(axial ratio of dark halo: $Q\sim0.3-0.4$).
However, these models assumed that the velocity dispersion of stars, $\overline{v^2_R}$ and $\overline{v^2_z}$ in the $R$ and $z$ directions in cylindrical coordinates, respectively, are identical.
That is, a velocity anisotropy parameter defined as $\beta_z=1-\overline{v^2_z}/\overline{v^2_R}$ is zero. Since there is a strong degeneracy between this anisotropy and dark halo shape as shown by \citet{2008MNRAS.390...71C}, considering a non-zero $\beta_z$ is needed to establish more reliable non-spherical dynamical models.
In turn, \citet{2015ApJ...810...22H} have revised axisymmetric mass models by taking into account a non-zero $\beta_z$ and revisited the shapes of the dark haloes in the MW dSphs. They found that these galaxies prefer to have elongated dark haloes, even introducing the effect of this velocity anisotropy of stars.
On the basis of \citet{2015ApJ...810...22H}, these models were applied to the most recent velocity data for not only classical but also ultra-faint dwarf galaxies~\citep{2016MNRAS.461.2914H} and to the multiple stellar components of Carina~\citep{2018MNRAS.481..250H}.
Inspired by the above studies, in this work we newly develop the non-spherical mass models based on a ULA dark halo profile predicted by numerical simulations.
More precisely, we solve axisymmetric Jeans equations and apply them to the kinematic data of the eight classical dSphs: Draco, Ursa Minor, Carina, Sextans, Leo I, Leo II, Sculptor and Fornax in order to investigate the global shape of ULA dark halo as well as to constrain on particle mass of ULA dark matter.

This paper is organized as follows.
In section~2, we explain the axisymmetric models for density profiles of stellar and ULA dark halo components based on an axisymmetric Jeans analysis, and briefly describe the observed data of the eight dSphs and the fitting procedure using unbinned likelihood functions.
In section~3, we present the results of the fitting analysis.
In section~4, we compare with other works, and discuss some extended models and implications for the origin of non-spherical ULA dark haloes.
Finally, we summarize our findings in Section~5.

\section{Models and Jeans analysis}
\subsection{Axisymmetric Jeans equations}
In an attempt to set the constraints on particle mass of ULA dark matter $m_{\psi}$ by dSphs,
we employ the Jeans analysis, especially using the axisymmetric mass models based on the axisymmetric Jeans equations.

Assuming that the system is settled in a dynamical equilibrium and collisionless under a smooth gravitational potential, spacial and velocity distributions of their stars are fundamentally described by the distribution function, which obeys the steady-state collisionless Boltzmann equation \citep{2008gady.book.....B}.
However, the current observations for the dSphs are able to resolve only the projected information of their internal phase-space distributions, such as the position of sky plane and line-of-sight velocity.
Thus, solving this equation by observations is not straightforward, so a commonly used approach is to take moments of this equation.
This is because the lower-order moments can be easily measured by observations, and thus we can compare the observable moments with those from the theoretical equation.
The equations taking moments of steady-state collisionless Boltzmann equation are so-called the Jeans equations.

In axisymmetric cases, the second-order axisymmetric Jeans equations are described as
\begin{eqnarray}
\overline{v^2_z} &=&  \frac{1}{\nu(R,z)}\int^{\infty}_z \nu\frac{\partial \Phi}{\partial z}dz,
\label{AGEb03}\\
\overline{v^2_{\phi}} &=& \frac{1}{1-\beta_z} \Biggl[ \overline{v^2_z} + \frac{R}{\nu}\frac{\partial(\nu\overline{v^2_z})}{\partial R} \Biggr] +
R \frac{\partial \Phi}{\partial R},
\label{AGEb04}
\end{eqnarray}
where $\nu$ is the three-dimensional stellar density and $\Phi$ is the gravitational potential dominated by dark matter, which means that stellar motions in a system are assumed to be obeyed by a dark matter potential only\footnote{Although the stellar motion in Fornax dSph could be affected by the stellar potentials themselves, because their dynamical mass to light ratio is not so much higher than unity~\citep{2012AJ....144....4M}, this is generally ignored and it is beyond the scope of this present work to take into account the effects of stellar potential on Jeans analysis.}.
We assume that the mixed moments such as $\overline{v_Rv_z}$ vanish and the velocity ellipsoid is aligned with the cylindrical coordinate.
We also assume that the density of tracer stars has the same orientation and symmetry as that of dark halo.
$\beta_z=1-\overline{v^2_z}/\overline{v^2_R}$ is a velocity anisotropy parameter introduced by~\citet{2008MNRAS.390...71C}.
In this work, $\beta_z$ is assumed to be constant for the sake of simplicity.
Nevertheless, this assumption is roughly consistent with dark matter simulations reported by~\citet{2014MNRAS.439.2863V} who have shown that simulated subhaloes have an almost constant $\beta_z$ or a weak trend as a function of radius along each axial direction.
Although the velocity second moments from equation~(\ref{AGEb03}) and (\ref{AGEb04}) are in principle provided by the second moments that separate into the contribution of random and ordered motions, as defined by $\overline{v^2}= \sigma^2+\overline{v}^2$, we assume that a dSph does not rotate and thus the second moment is comparable to the velocity dispersion.
This is because
the line-of-sight velocity gradients of dSphs are so tiny and these gradients can be explained by projection effects~\citep[e.g.,][]{2008ApJ...688L..75W,2008ApJ...681L..13B,2011MNRAS.411.1013B}.
Even if a rotational velocity exists in the dSphs, a ratio of the velocity to dispersion is considerably small, thereby implying that dSphs are dispersion-supported stellar systems~\citep[e.g.,][]{2017MNRAS.465.2420W}

\begin{table*}
	\centering
	\caption{The observational dataset for MW dSph satellites.}
	\label{table1}
	\begin{tabular}{rrrrrrrrr} 
		\hline\hline
Object & $N_{\rm sample}$ & RA(J2000) & DEC(J2000)   & $D_{\odot}$ & $b_{\ast}$ &  $q^{\prime}$ & $\langle u\rangle_{obs}$ &  $r_{90}$\\
            &                                 & [hh:mm:ss]   &  [dd:mm:ss]     &      [kpc]         &        [pc]              &    (axial ratio) & [km~s$^{-1}$] & [pc]\\
		\hline
		Draco         & 468$^{(9)}$    & 17:20:12.4   & $+$57:54:55     & $ 76\pm 6^{(2)}$  & $214\pm  2^{(1)}$  & $0.71\pm0.01^{(1)}$ & $-290.0^{(9)}$ & 526$^{(9)}$ \\
		Ursa~Minor    & 313$^{(10)}$    & 15:08:08.5   & $+$67:13:21     & $ 76\pm 3^{(3)}$  & $407\pm  2^{(1)}$  & $0.45\pm0.01^{(1)}$ & $-246.9^{(10)}$ & 559$^{(10)}$  \\
		Carina        & 1086$^{(11)}$   & 06:41:36.7   & $-$50:57:58     & $106\pm 6^{(4)}$  & $308\pm  23^{(1)}$  & $0.64\pm0.01^{(1)}$ & $220.7^{(11)}$ & 428$^{(11)}$ \\
		Sextans       & 445$^{(12)}$    & 10:13:03.0   & $-$01:36:53     & $ 86\pm 4^{(5)}$  & $413\pm  3^{(1)}$  & $0.70\pm0.01^{(1)}$ & $224.3^{(12)}$ & 716$^{(12)}$ \\
		Leo~I         & 328$^{(13)}$    & 10:08:28.1   & $+$12:18:23     & $254\pm15^{(6)}$  & $270\pm  2^{(1)}$  & $0.70\pm0.01^{(1)}$ & $282.9^{(13)}$ & 496$^{(13)}$ \\
		Leo~II        & 177$^{(14)}$    & 11:13:28.8   & $+$22:09:06     & $233\pm14^{(7)}$  & $171\pm  2^{(1)}$  & $0.93\pm0.01^{(1)}$ & $ 78.7^{(14)}$ & 351$^{(14)}$ \\
		Sculptor      & 1360$^{(12)}$   & 01:00:09.4   & $-$33:42:33     & $ 86\pm 6^{(8)}$  & $280\pm  1^{(1)}$  & $0.67\pm0.01^{(1)}$ & $111.4^{(12)}$ & 468$^{(12)}$ \\
		Fornax        & 2523$^{(12)}$   & 02:39:59.3   & $-$34:26:57     & $147\pm12^{(4)}$  & $838\pm  3^{(1)}$  & $0.71\pm0.01^{(1)}$ & $ 55.2^{(12)}$ & 983$^{(12)}$ \\
	\hline
	\end{tabular}
\begin{flushleft}
References: (1)~\citet{2018ApJ...860...66M};  
(2)~\citet{2004AJ....127..861B}; (3)~\citet{2002AJ....123.3199C}; (4)~\citet{2009AJ....138..459P}; (5)~\citet{2009ApJ...703..692L}; (6)~\citet{2004MNRAS.354..708B}; (7)~\citet{2005MNRAS.360..185B}; (8)~\citet{2008AJ....135.1993P}; (9)~\citet{2015MNRAS.448.2717W}; (10)~\citet{2018AJ....156..257S}; (11)~\citet{2016ApJ...830..126F}; (12)~\citet{2009AJ....137.3100W}; (13)~\citet{2008ApJ...675..201M}; (14)~\citet{2007AJ....134..566K};
\end{flushleft}
\end{table*}

In order to compare with the observable velocity second moments of the dSphs, we ought to integrate the intrinsic velocity second moments~$(\overline{v^2_R}(=(1-\beta_z)^{-1}\overline{v^2_z}),\overline{v^2_\phi},\overline{v^2_z})$ along the line-of-sight, considering the inclination of the dSph with respect to the observer. 
To this end, we project the intrinsic velocity second moments in two steps followed by some previous works~\citep{1997MNRAS.287...35R,2006MNRAS.371.1269T,2012ApJ...755..145H}.
Firstly, we project $\overline{v^2_R}$ and $\overline{v^2_{\phi}}$ to the plane parallel to the galactic plane along the intrinsic major axis.
This dispersion is described as
\begin{equation}
\overline{v^2_{\ast}} = \overline{v^2_{\phi}}\frac{x^2}{R^2} + \overline{v^2_R}\Bigl(1-\frac{x^2}{R^2}\Bigr),
\label{los1}
\end{equation}
where $x$ is the projected coordinate on the sky plane.
Secondly, assigning the angle $\theta$ between the line of sight and the galactic plane~($\theta=90^{\circ}-i$, which $i$ is an inclination angle explained below), we derive the line-of-sight velocity second moments using $\overline{v^2_z}$ and $\overline{v^2_{\ast}}$,
\begin{equation}
\overline{v^2_{\ell}} = \overline{v^2_{\ast}}\cos^2\theta + \overline{v^2_z}\sin^2\theta.
\label{los2}
\end{equation}
We compute the luminosity-weighted average of $\overline{v^2_{\ell}}$ along the line of sight, so as to compare the theoretical velocity second moments with the observed ones.
This quantity can be written as 
\begin{equation}
\overline{v^2_{\rm l.o.s}}(x,y) = \frac{1}{I(x,y)}\int^{\infty}_{-\infty}\nu(R,z)\overline{v^2_{\ell}}(R,z)d\ell,
\label{los3}
\end{equation}
where $I(x,y)$ indicates the surface stellar density profile calculated from $\nu(R,z)$, and $(x,y)$ are the sky coordinates aligned with the major and minor axes, respectively.

For stellar density distributions of the dSphs, we assume {\it oblate}\footnote{Here, we suppose that the stellar distribution of dSph has an oblate shape only, because \citet{2015ApJ...810...22H} concluded that most of stellar distributions are much better fitted by the oblate shape than by the prolate ones.} Plummer profile~\citep{1911MNRAS..71..460P} generalized to an axisymmetric form:
$\nu(R,z)=(3L/4\pi b^3_{\ast})(1+m^2_{\ast}/b^2_{\ast})^{-5/2}$, where $m^2_{\ast}=R^2+z^2/q^2$, so that $\nu$ is constant on spheroidal shells with axial ratio $q$, and $L$ and $b_{\ast}$ are the total luminosity and the half-light radius along the major axis, respectively.
This form can be analytically calculated from the surface density using Abel transformation, $I(x,y)=(L/\pi b^2_{\ast})(1+m^{\prime 2}_{\ast}/b^2_{\ast})^{-2}$,
where $m^{\prime 2}_{\ast}=x^2+y^2/q^{\prime 2}$.
The projected axial ratio $q^{\prime}$ is related to the intrinsic one $q$ and the inclination angle $i$~$(=90^{\circ}-\theta)$, such as $q^{\prime 2} = \cos^2i+q^2\sin^2i$.
Since this relationship can be rewritten as $q=\sqrt{q^{\prime 2}-\cos^2i}/\sin i$, the domain of inclination angle is bounded within the range of $0\leq\cos^2i<q^{\prime 2}$.

\subsection{ULA dark matter dark matter density profile}
Using high resolution simulations for ULA dark matter, \citet[][hereafter S14]{2014NatPh..10..496S} found that the structure of a dark matter halo is characterized by a soliton core at the centre of the halo and an asymptotic~Navarro-Frenk-White~\citep[hereafter NFW,][]{1997ApJ...490..493N} profile beyond the central soliton core.

The soliton core can be created in each ULA dark matter halo and be described by the ground state solution of the Schr\"odinger-Poisson equations.
S14 also found that the central region can be well fitted by
\begin{equation}
\rho_{\rm soliton}(r) = \frac{\rho_c}{\Bigl[1+0.091\Bigl(\frac{r}{r_c}\Bigr)^2\Bigr]^8},
\label{soliton}
\end{equation}
where $r_c$ is the soliton core radius and $\rho_c$ is the central density given by
\begin{equation}
\rho_c = 1.9\times10^{12} \Bigl(\frac{m_{\psi}}{10^{-23}{\rm eV}}\Bigr)^{-2}\Bigl(\frac{r_c}{\rm pc}\Bigr)^{-4} \hspace{1.3mm} [M_{\odot}{\rm pc}^{-3}],
\label{solitoncore}
\end{equation}
with ULA dark matter mass, $m_{\psi}$.
The soliton core radius is characterized by the Broglie wavelength $r_c\sim\lambda_{\text dB}\equiv \hbar/(m_\psi v)$ ($\hbar$ and $v$ are the reduced Planck constant and the velocity of particle, respectively) and typical physical size for ULA dark matter corresponds to several~kpc~$\lambda_{\text dB}=0.1-1$~kpc with $m_{\psi} = 10^{-22}-10^{-23}\text{eV}$.

On the other hand, an outer envelope of the dark matter density profile is fitted by an NFW profile,
\begin{equation}
\rho_{\rm NFW}(r) = \frac{\rho_{s}}{(r/r_s)(1+r/r_s)^2},
\label{nfw}
\end{equation}
where $\rho_{s}$ and $r_s$ are a scale density and a scale radius, respectively.
In order to sustain a continuity of both density profiles, we introduce the transition radius $r_{\epsilon}$ between the inner soliton core and the outer NFW profile.
Using density ratio $\epsilon$, this relation can be written as 
\begin{equation}
\dfrac{\rho_{s}}{(r_\epsilon/r_s)(1+r_\epsilon/r_s)^2}=\dfrac{\rho_{c}}{[1+0.091(r_\epsilon/r_c)^2]^8} = \epsilon\rho_c,
\end{equation}
where $r_{\epsilon}$ reads 
\begin{equation}
r_\epsilon=\dfrac{r_c}{\sqrt{0.091}}(\epsilon^{-1/8}-1)^{1/2} \ .
\label{Rtran}
\end{equation}  
Therefore, the total density profile of ULA dark matter is
\begin{eqnarray}
\rho(r) = \begin{cases}
\dfrac{\rho_c}{[1+0.091(r/r_c)^2]^8} \quad (r < r_\epsilon) \\
\dfrac{\rho_{s}}{(r/r_s)(1+r/r_s)^2} \quad (r \geq r_\epsilon) \ .
\end{cases}
\label{solNFW}
\end{eqnarray}
 In this work, we investigate non-spherical, especially axisymmetric dark matter distributions of the dSphs.
Hence $r$ is transformed from spherical to axisymmetric form: 
\begin{equation}
r^2 = R^2 + z^2/Q^2,
\end{equation} 
where $Q$ denotes the axial ratio of dark matter halo.

According to the results from numerical simulations~\citep[e.g.,][]{2014NatPh..10..496S,2014PhRvL.113z1302S,2017MNRAS.471.4559M},
the transition radius is generally greater than $3r_c$, which corresponds to several kpc.
A typical value of $r_c$ on dSph mass scales at redshift zero can be estimated as
\begin{equation}
r_c={\rm 16.0\ kpc}\ \Bigl(\frac{M_{\rm halo}}{10^9M_{\odot}}\Bigr)^{-1/3} \Bigl(\frac{m_{\psi}}{10^{-23}{\rm eV}}\Bigr)^{-1},
\end{equation}
where $M_{\rm halo}$ is a virial mass of dark matter halo~(see S14).
When we consider a typical dark matter halo of a dSph~($M_{\rm star}\sim10^6 M_{\odot}$), we can naively estimate $M_{\rm halo}\sim10^9M_{\odot}$ due to the stellar-to-dark matter halo mass relation~\citep[e.g.,][]{2015MNRAS.448.2941S}, and the resultant core radius should be $r_c\sim1-10$~kpc with $m_{\psi} = 10^{-22}-10^{-23}\text{eV}$. 
On the other hand, the observed half-light radii~$(b_{\ast})$ and the radius that encloses 90~per~cent of the spectroscopic sample~$(r_{90})$ of the dSphs are smaller than 1~kpc~(shown in Table~\ref{table1}).
Therefore, the transition radius is a few times greater than these radii, and thus most of observed member stars can be settled within the central soliton core.
This means it is reasonable to suppose that motions of all stars are governed by the gravitational potential stemmed from the central soliton core only and ignore the outer NFW halo in the first place when implementing the Jeans analysis.
Nevertheless, in order to investigate whether this assumption is justified or not, we also extend the above analysis to include outer NFW halo potential and have a discussion on those results~(see Section~4.2).

\subsection{The Galactic Dwarf Spheroidal Galaxy Data}
In this work, we adopt the photometric and spectroscopic data of the resolved member stars in the eight luminous dSphs in the Milky Way: Draco, Ursa~Minor, Carina, Sextans, Leo~I, Leo~II, Sculptor, and Fornax.
This is because each have $\gtrsim200$~stars with line-of-sight velocities from spectroscopy and well-constrained stellar distributions from photometry, so that these sample volumes might be enough to obtain more stringent limits on the mass range of $m_\psi$ than those from other fainter dSphs such as ultra faint dwarf galaxies.
Moreover, these galaxies have larger velocity dispersions~($\gtrsim10$~km~s$^{-1}$), so that the influences of unresolved binary stars on the velocity dispersion measurements of each galaxy will be negligible~\citep{2010ApJ...721.1142M,2013ApJ...779..116M,2017ApJ...836..202S,2018AJ....156..257S}.

The observed properties of these eight dSphs are tabulated in Table~\ref{table1}:~the number of kinematic sample stars, the central sky coordinates, distances, projected half-light radii, projected stellar axial ratios, mean velocities, and the radius that encloses 90~per~cent of the spectroscopic sample to compare with a transition radius of ULA dark matter halo. 
These observed values are adopted from each observational paper listed in Table~\ref{table1}.

For the stellar velocity samples of their member stars, we adopt the latest data published by~\citet{2016ApJ...830..126F} for Carina, by \citet{2015MNRAS.448.2717W} for Draco,
by \citet{2018AJ....156..257S} for  Ursa~Minor, by~\citet[][]{2008ApJ...675..201M} for Leo~I, by~\citet{2007AJ....134..566K} for Leo~II, and by \citet{2009AJ....137.3100W,2009AJ....137.3109W} for Fornax, Sculptor and Sextans, respectively.

\begin{table*}
	\centering
	\caption{Parameter constraints for MW dSph satellites. Errors correspond to the $1\sigma$ range of our analysis.}
	\label{table2}
	\scalebox{1.2}[1.2]{
	\begin{tabular}{ccccccc} 
		\hline\hline
Object    & $Q$ &   $\log_{10}(r_{c})$  & $\log_{10}$($m_{\psi})$       & $-\log_{10}(1-\beta_z)$   &$i$  & $\log_{10}(\rho_{c})$ \\
      &   &  [pc]  & [$10^{-23}{\text eV}$]       &  &[deg]  & [$M_{\odot}$~kpc$^{-3}$] \\
		\hline
Draco             & $0.22^{+0.08}_{-0.05}$  & $2.88^{+0.07}_{-0.06}$ & $0.59^{+0.13}_{-0.14}$  & $-0.19^{+0.09}_{-0.08}$ &  $77.08^{+8.28}_{-11.32}$  & $ 8.57 ^{+ 0.10 } _{- 0.11 } $\\
Ursa~Minor        & $1.11^{+0.48}_{-0.44}$  & $2.50^{+0.14}_{-0.13}$ & $1.55^{+0.18}_{-0.25}$  & $ 0.55^{+0.16}_{-0.11}$ &  $79.18^{+6.92}_{ -7.65}$    & $ 8.23 ^{+ 0.19 } _{- 0.17 } $\\
Carina            & $0.30^{+0.25}_{-0.11}$  & $2.90^{+0.15}_{-0.13}$ & $0.82^{+0.27}_{-0.29}$  & $-0.01^{+0.22}_{-0.15}$ &  $73.31^{+10.94}_{-11.13}$ &$ 8.03 ^{+ 0.16 } _{- 0.18 } $\\
Sextans           & $0.77^{+0.62}_{-0.40}$  & $2.74^{+0.17}_{-0.15}$ & $1.22^{+0.28}_{-0.33}$  & $ 0.16^{+0.20}_{-0.17}$ &  $69.80^{+13.42}_{-12.20}$  & $ 7.88 ^{+ 0.19 } _{- 0.16 } $\\
Leo~I             & $0.44^{+0.32}_{-0.17}$  & $2.45^{+0.11}_{-0.11}$ & $1.38^{+0.20}_{-0.19}$  & $-0.04^{+0.14}_{-0.09}$ &  $70.07^{+12.75}_{-13.33}$ & $ 8.70 ^{+ 0.22 } _{- 0.19 } $\\
Leo~II            & $1.13^{+0.56}_{-0.62}$  & $2.61^{+0.82}_{-0.25}$ & $1.29^{+0.42}_{-1.51}$  & $ 0.22^{+0.21}_{-0.25}$ &  $59.48^{+19.26}_{-15.12}$  & $ 8.25 ^{+ 0.22 } _{- 0.25 } $\\
Sculptor          & $0.29^{+0.16}_{-0.09}$  & $2.69^{+0.06}_{-0.06}$ & $0.92^{+0.20}_{-0.14}$  & $ 0.18^{+0.16}_{-0.16}$ &  $53.93^{+19.55}_{ -2.20}$  & $ 8.72 ^{+ 0.15 } _{- 0.38 } $\\
Fornax            & $0.75^{+0.35}_{-0.25}$  & $2.92^{+0.09}_{-0.09}$ & $0.92^{+0.17}_{-0.19}$  & $ 0.19^{+0.13}_{-0.10}$ &  $69.17^{+13.55}_{-11.40}$  & $ 7.76 ^{+ 0.13 } _{- 0.10 } $\\
	\hline
	\end{tabular}
	}
\end{table*}

\subsection{Fitting Procedure}
Our aim is to obtain the dark matter halo structural parameters, especially ULA dark matter mass $m_{\psi}$ by fitting our non-spherical mass models to the velocity second moments of each dSph.
The fitting procedure in this work is different from those in some previous works~\citep[e.g.,][]{2017MNRAS.468.1338C,2017MNRAS.472.1346G}. \citet{2017MNRAS.468.1338C} adopted the method of fitting their dynamical mass models to the line-of-sight velocity second moment profiles built from the individual stellar velocities of dSphs through a likelihood function \citep[see equation (5) in][]{2017MNRAS.468.1338C}.
On the other hand, our work adopts the Gaussian distribution of the line-of-sight velocity to compare the observed and theoretical velocity second moments.
To do this, we assume that the line-of-sight velocity distribution is Gaussian and centred on the systemic velocity
of the galaxy $\langle u \rangle$.
Given that the total number of tracers for each dSph is $N$, and the measured line-of-sight velocity of the
$i$th tracer and its observational error is defined by $u_i\pm\delta_{u,i}$ at the sky plane coordinates~$(x_i,y_i)$, the
likelihood function is described as
\begin{equation}
{\cal L} = \prod^{N}_{i=1}\frac{1}{(2\pi)^{1/2}[(\delta_{u,i})^2 + (\sigma_i)^2]^{1/2}}\exp\Bigl[-\frac{1}{2}\frac{(u_i-\langle u \rangle)^2}{(\delta_{u,i})^2 + (\sigma_i)^2} \Bigr],
\label{LF}
\end{equation}
where $\sigma_i$ is the theoretical line-of-sight velocity dispersion at~$(x_i,y_i)$ which is specified by model
parameters~(as described in the previous section) and derived from the Jeans equations.

The systemic velocity $\langle u \rangle$ of the dSph is a nuisance parameter that we marginalize over as a flat prior.
We impose that the parameter range of $\langle u \rangle$ is plus or minus 5~km~s$^{-1}$ from the observed mean velocities of each dSph (see the 8th column in Table~\ref{table1}).
Beside $\langle u \rangle$, we adopt uniform and Jeffreys priors to our five model parameters~$(Q,r_c,m_{\psi},\beta_z,i)$ over the following ranges:
\begin{enumerate}
\item $0.15\leq Q\leq2.0$;
\item $-2.0\leq \log_{10}[r_c/{\text pc}]\leq5.0$;
\item $-3.0\leq \log_{10}[m_{\psi}/10^{-23}{\text eV}]\leq3.0$;
\item $-1.0\leq -\log_{10}[1-\beta_{z}]<1.0$;
\item $\cos^{-1}(q^{\prime})< i/{\rm deg} \leq 90.0$;
\item $\langle u \rangle_{obs}-5~{\rm km}~{\rm s}^{-1}\leq\langle u \rangle\leq\langle u \rangle_{obs}+5~{\rm km}~{\rm s}^{-1}$.
\end{enumerate}
In order to get the posterior probability distribution function (PDF) of each parameter, we use a Markov Chain Monte Carlo~(MCMC) techniques, based on Bayesian parameter inference, with the Metropolis-Hastings algorithm\citep{1953JChPh..21.1087M,10.1093/biomet/57.1.97}.
Then, in order to avoid an influence of initial conditions and to generate independent samples, we take several post-processing steps such as burn-in step, the sampling step and length of the chain.
Using these PDFs, we calculate the percentiles of these PDFs to estimate credible intervals for each parameter straightforwardly\footnote{Following previous works, we fix the values of distance, half-light radius, and axial ratio of dSphs in this paper.}.

\begin{figure*}
	\begin{minipage}{0.49\hsize}
		\begin{center}
			\includegraphics[width=\columnwidth]{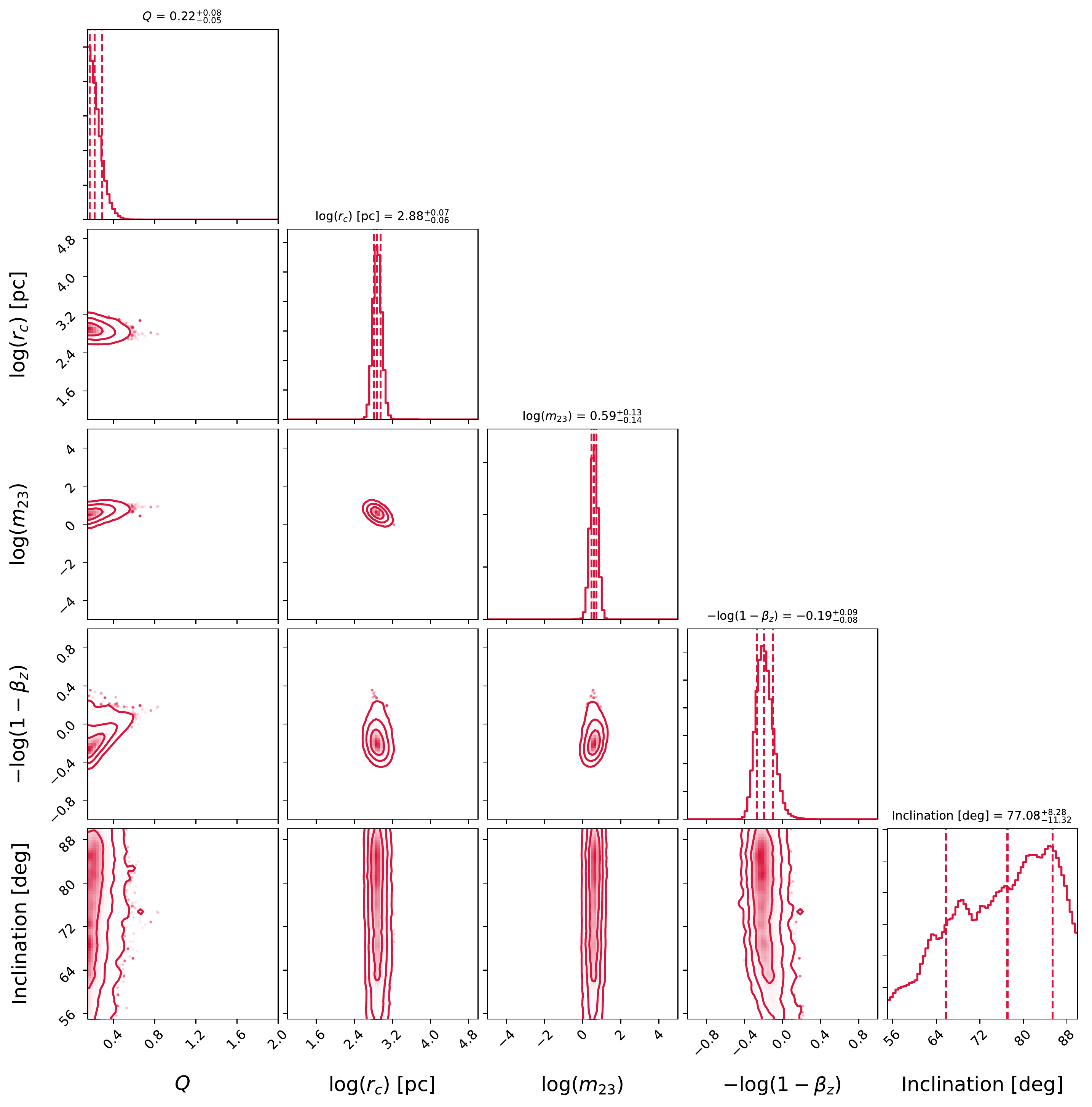}
		\end{center}
	\end{minipage}
	\begin{minipage}{0.49\hsize}
		\begin{center}
			\includegraphics[width=\columnwidth]{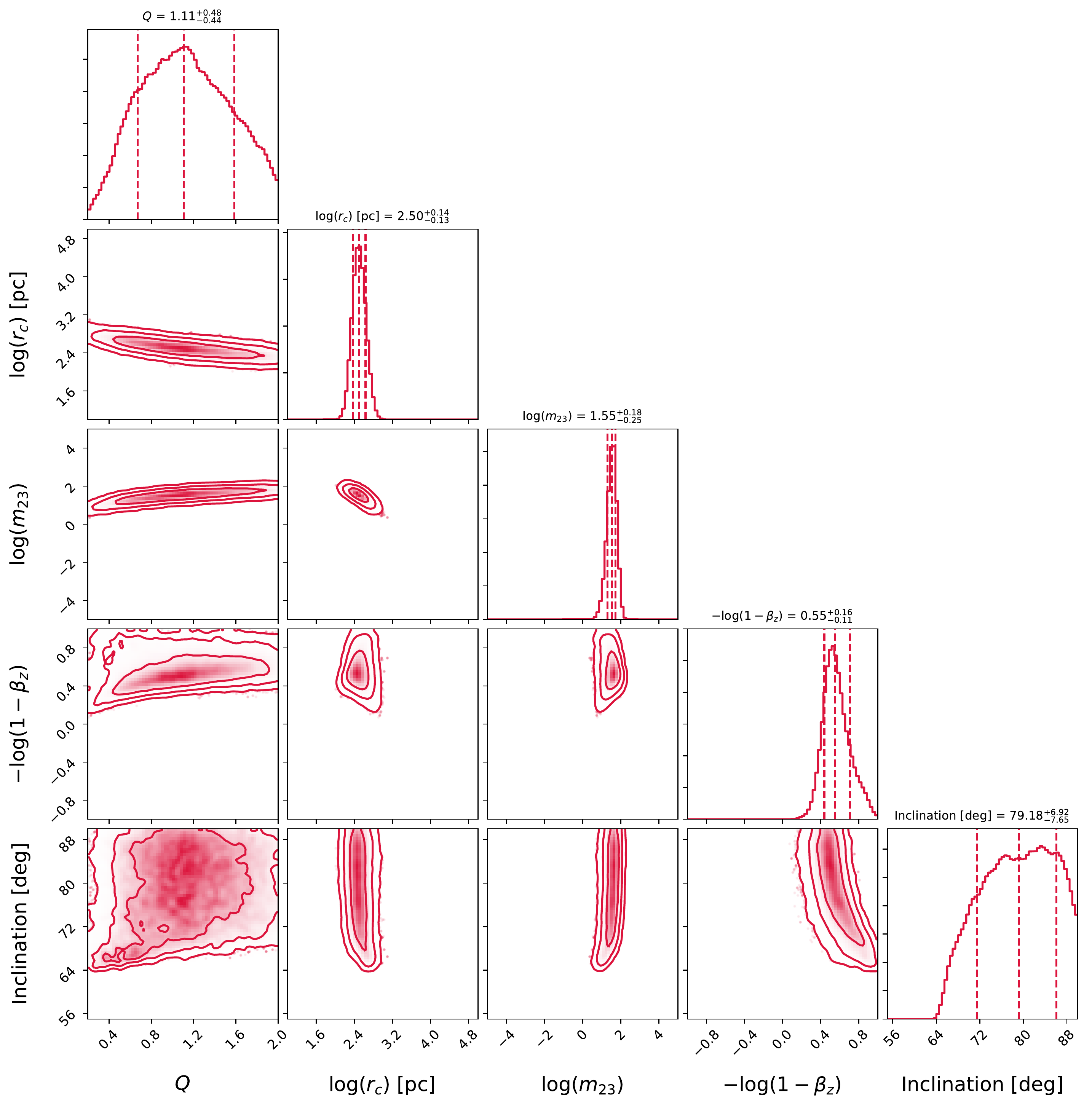}
		\end{center}
	\end{minipage}
    \caption{Posterior distribution functions of dark matter halo parameters for Draco~(left) and Ursa~Minor~(right). $m_{23}$ is particle mass of ULA dark matter $m_{\psi}$ normalized by $10^{-23}$~eV, that is, $m_{23}=m_{\psi}/10^{-23}$~eV.}
    \label{draumi}
\end{figure*}

\begin{figure*}
	\begin{minipage}{0.49\hsize}
		\begin{center}
			\includegraphics[width=\columnwidth]{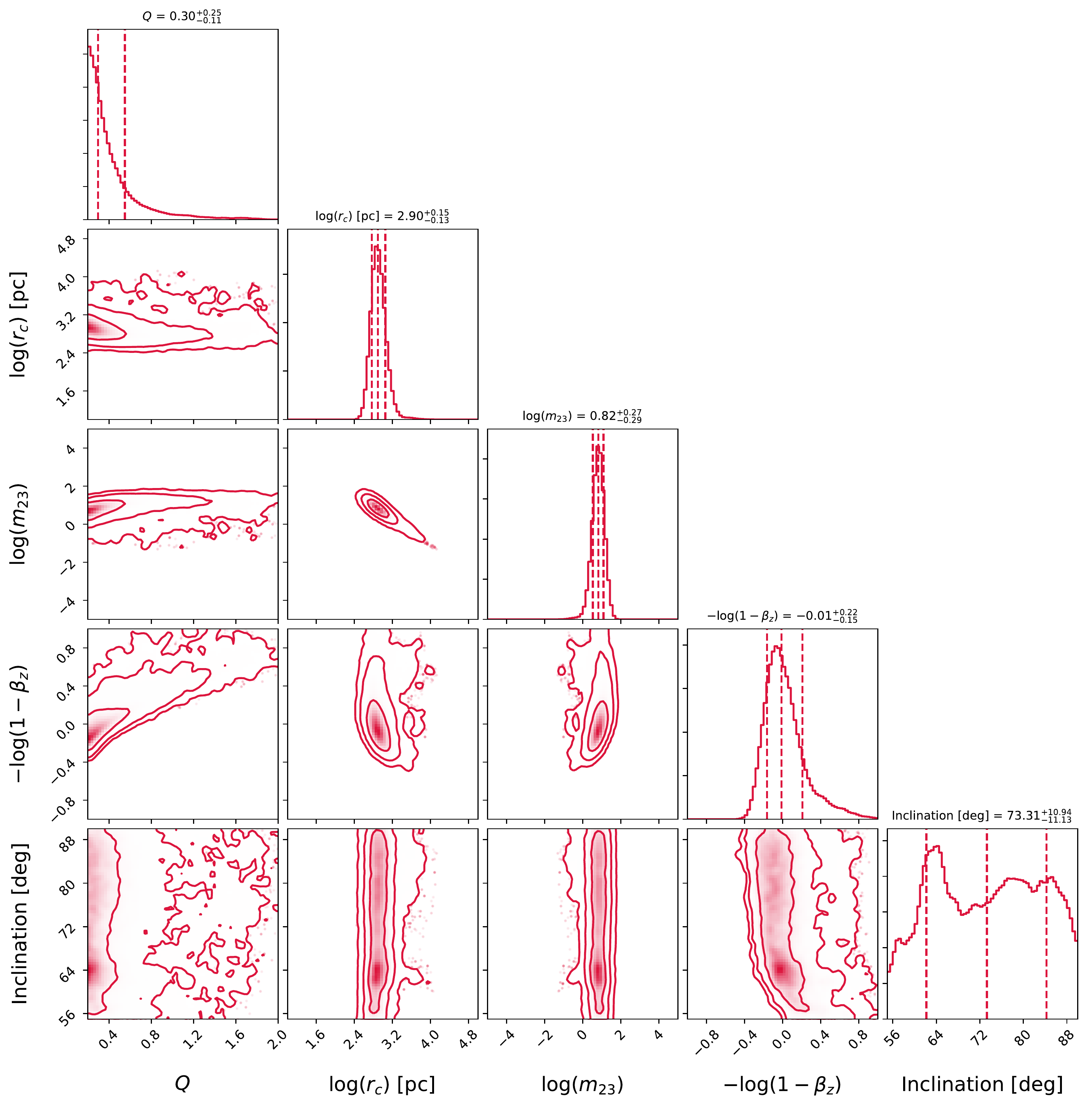}
		\end{center}
	\end{minipage}
	\begin{minipage}{0.49\hsize}
		\begin{center}
			\includegraphics[width=\columnwidth]{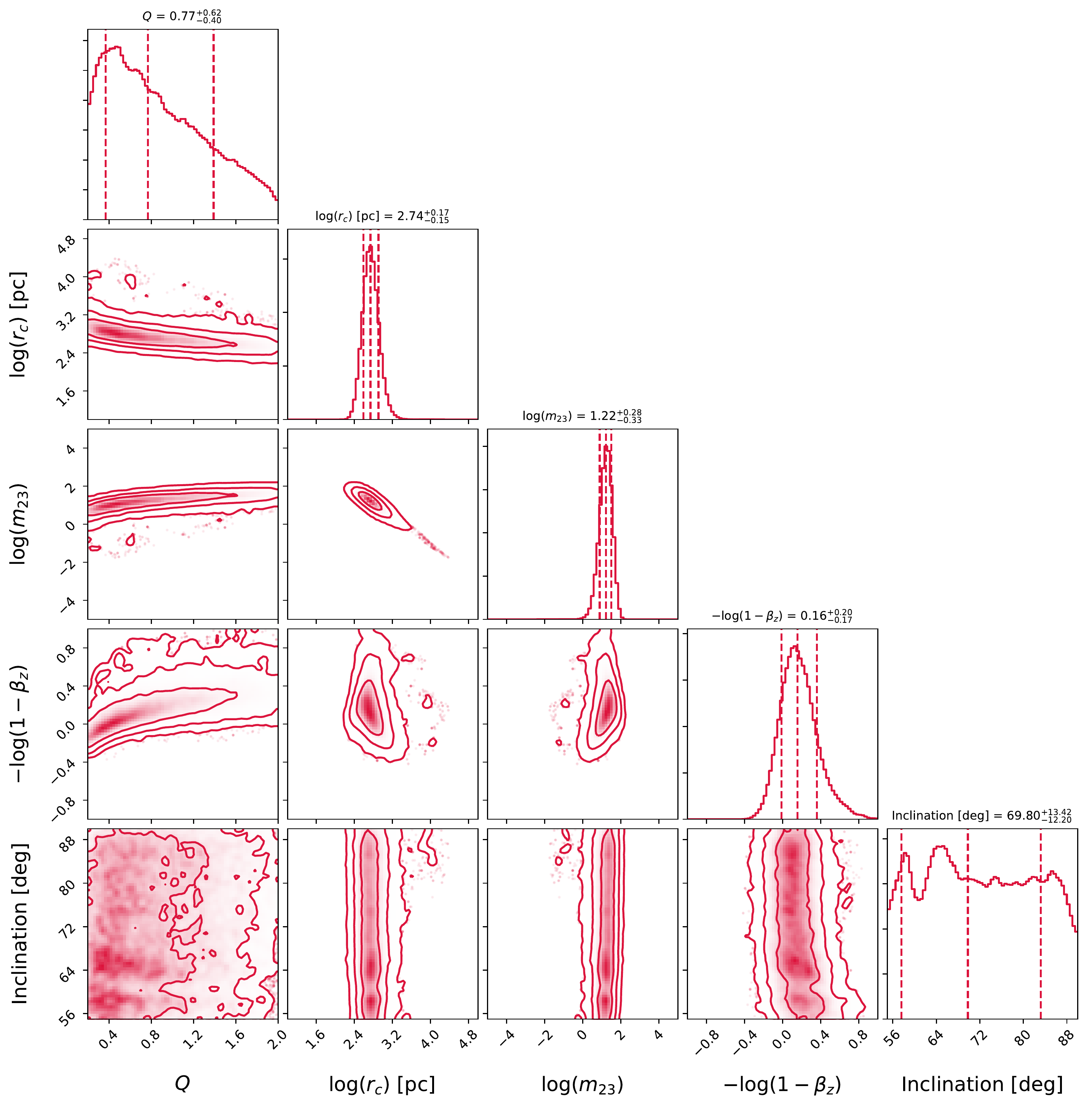}
		\end{center}
	\end{minipage}
    \caption{Same as figure~\ref{draumi}, but for Carina~(left) and Sextans~(right).}
    \label{carsex}
\end{figure*}

\begin{figure*}
	\begin{minipage}{0.49\hsize}
		\begin{center}
			\includegraphics[width=\columnwidth]{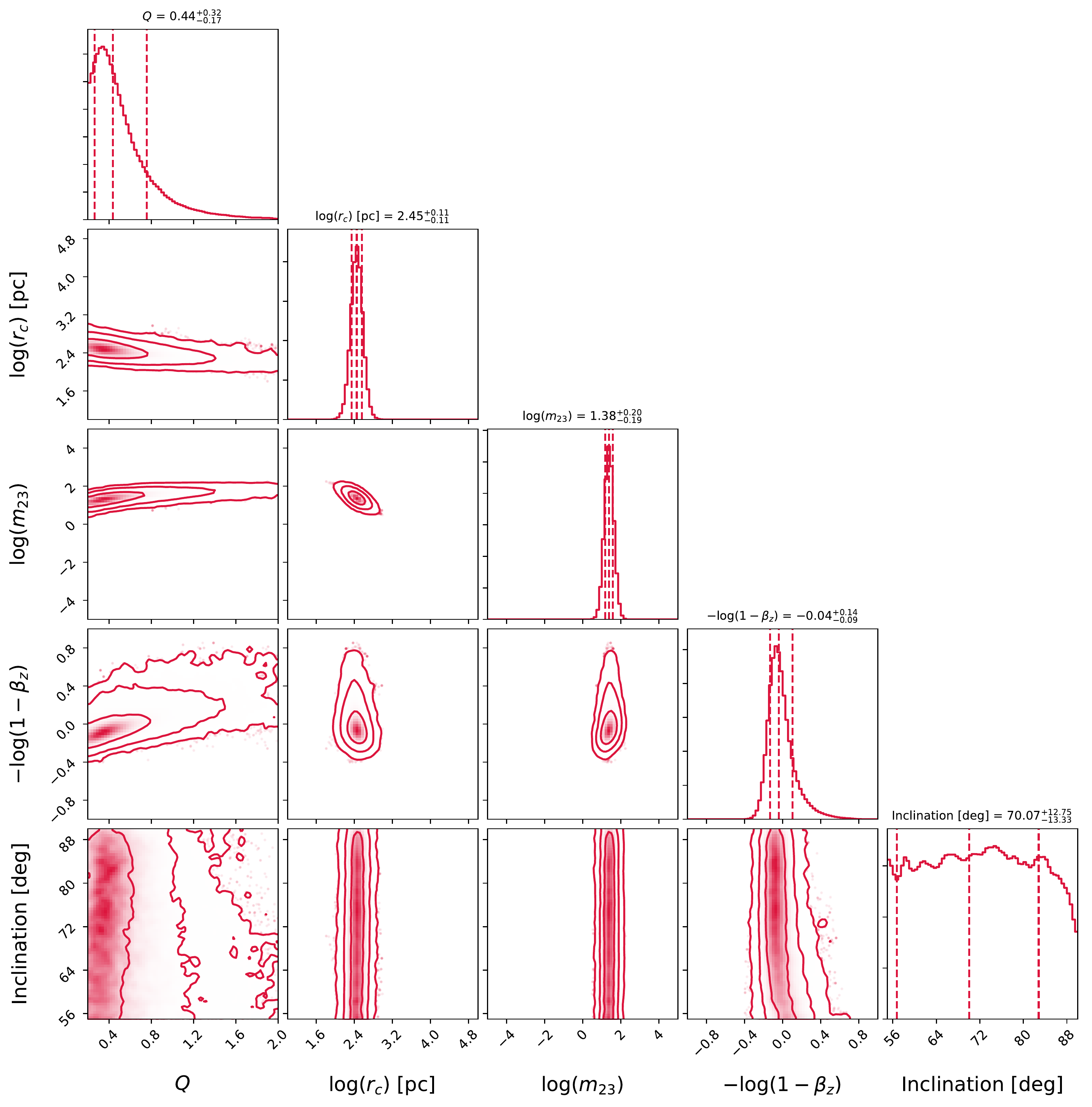}
		\end{center}
	\end{minipage}
	\begin{minipage}{0.49\hsize}
		\begin{center}
			\includegraphics[width=\columnwidth]{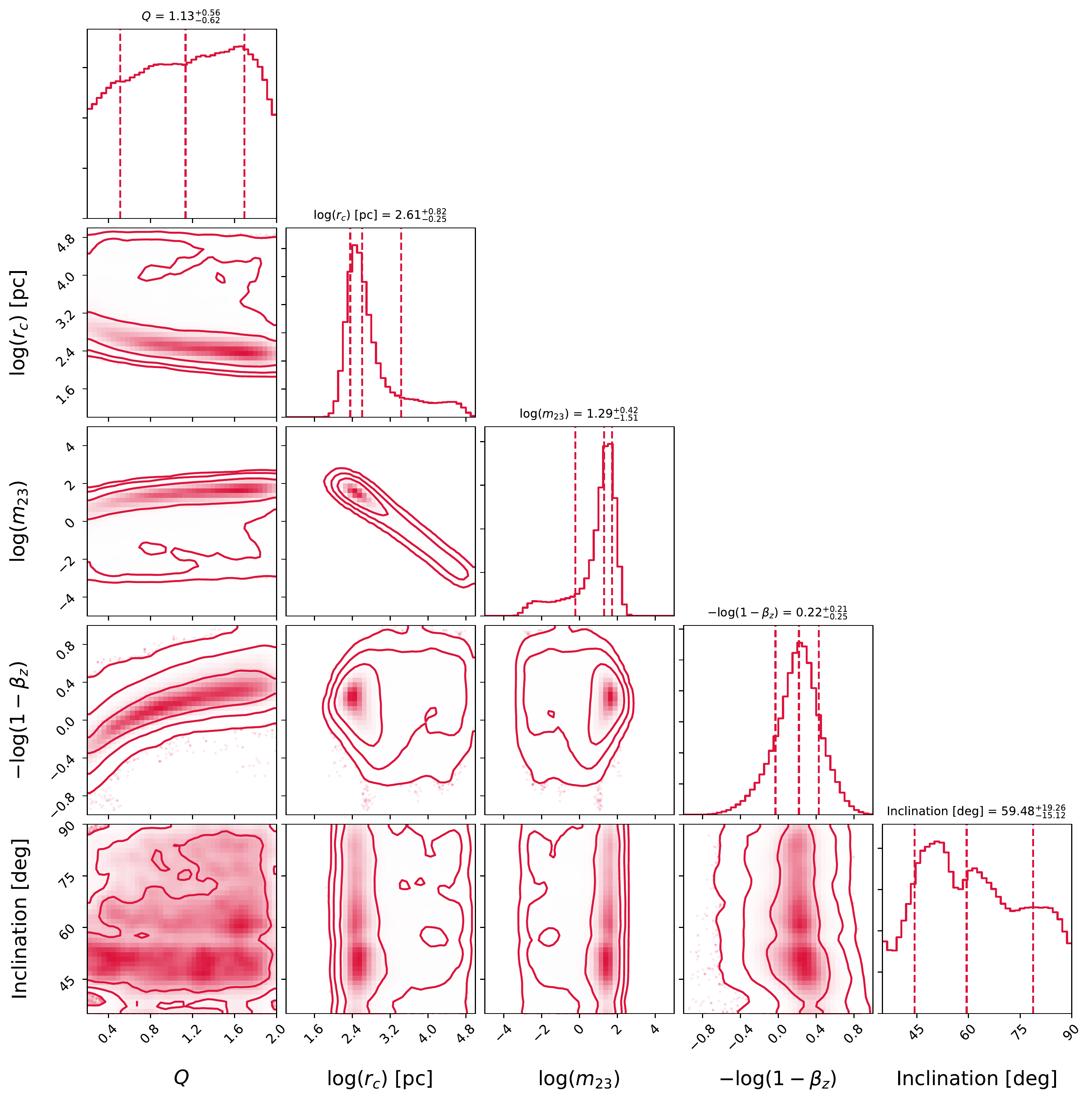}
		\end{center}
	\end{minipage}
    \caption{Same as figure~\ref{draumi}, but for Leo~I~(left) and Leo~II~(right).}
    \label{leo12}
\end{figure*}

\begin{figure*}
	\begin{minipage}{0.49\hsize}
		\begin{center}
			\includegraphics[width=\columnwidth]{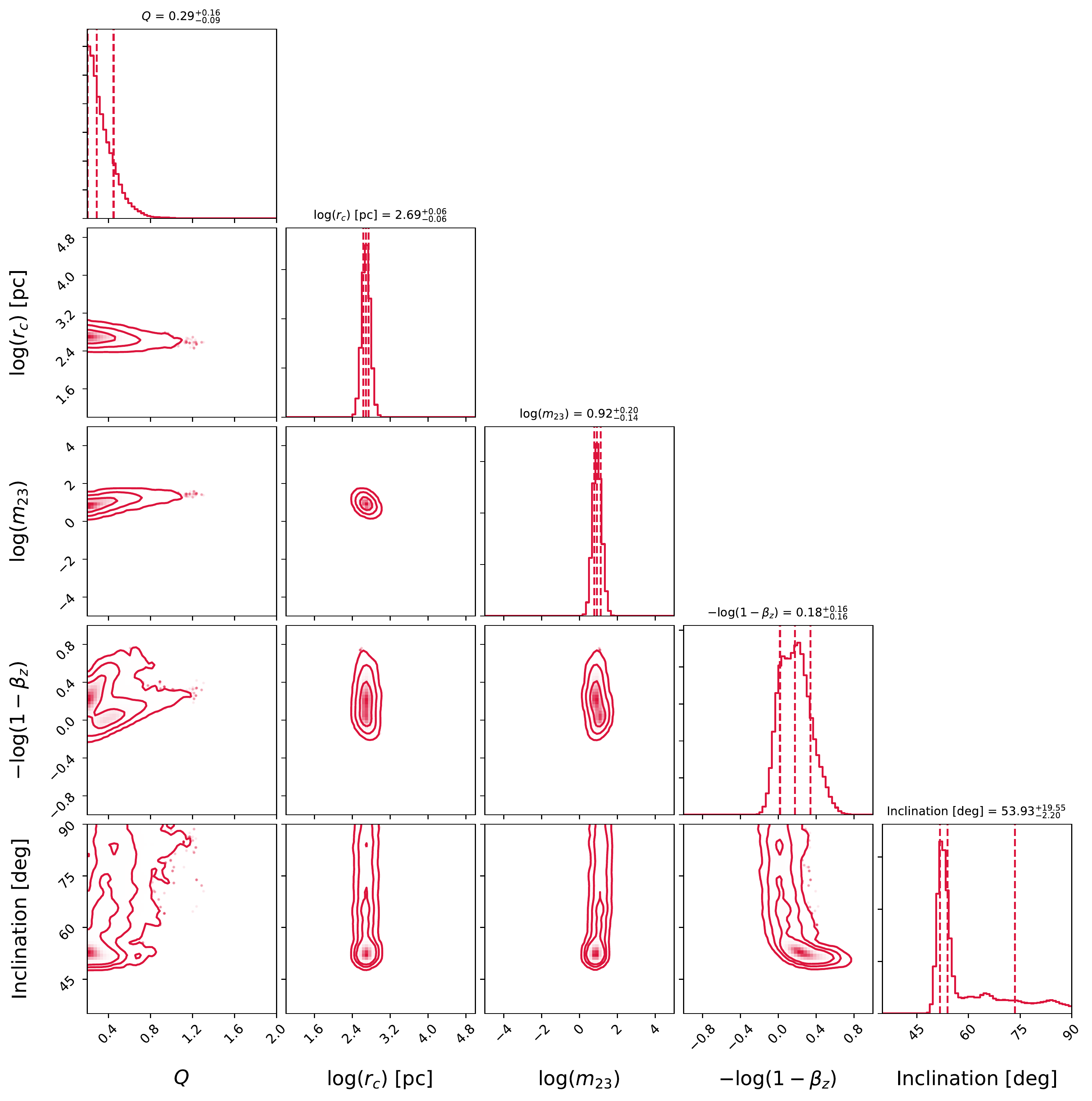}
		\end{center}
	\end{minipage}
	\begin{minipage}{0.49\hsize}
		\begin{center}
			\includegraphics[width=\columnwidth]{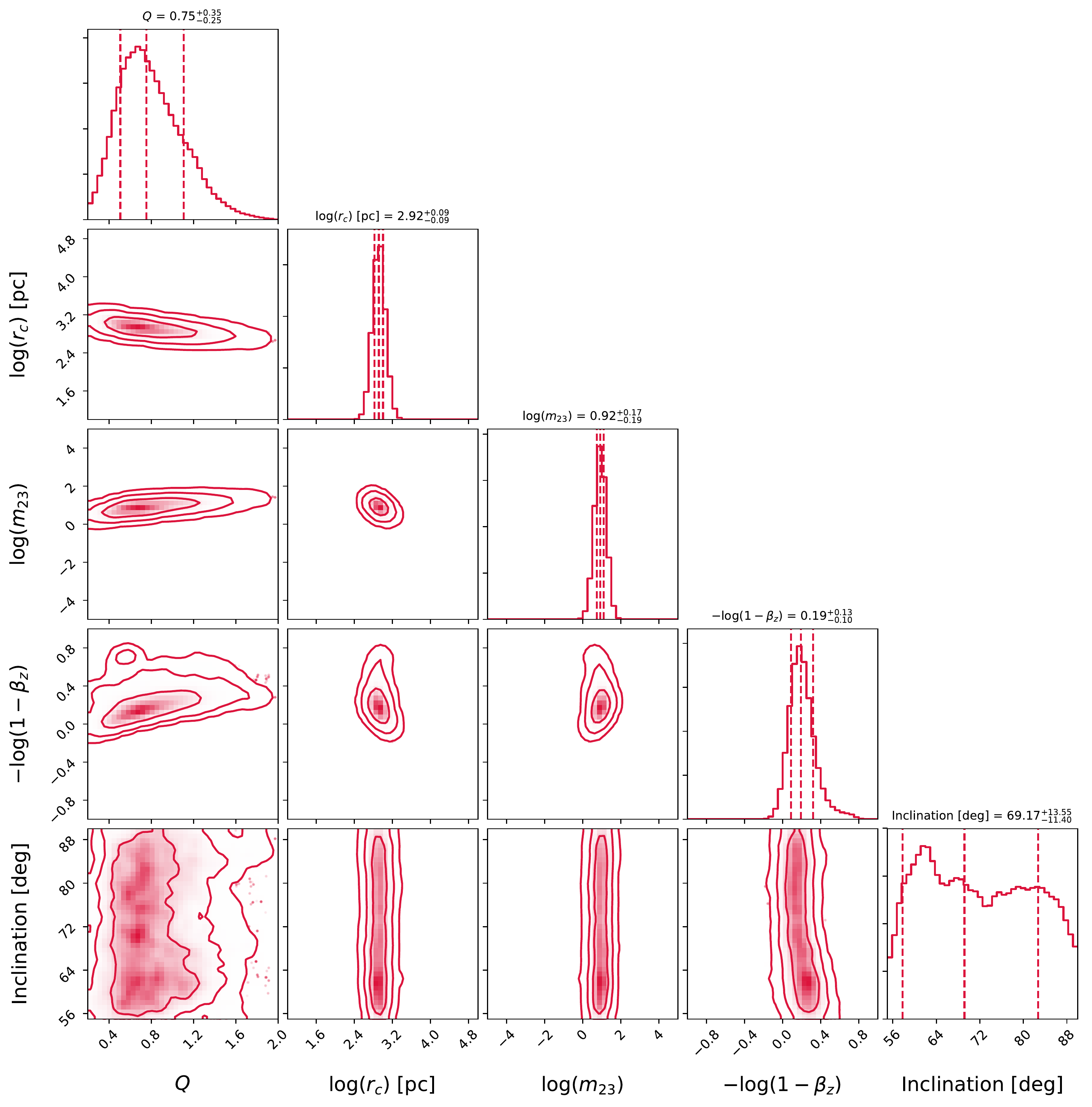}
		\end{center}
	\end{minipage}
    \caption{Same as figure~\ref{draumi}, but for Sculptor~(left) and Fornax~(right).}
    \label{sclfnx}
\end{figure*}

\begin{figure*}
	\includegraphics[scale=0.35]{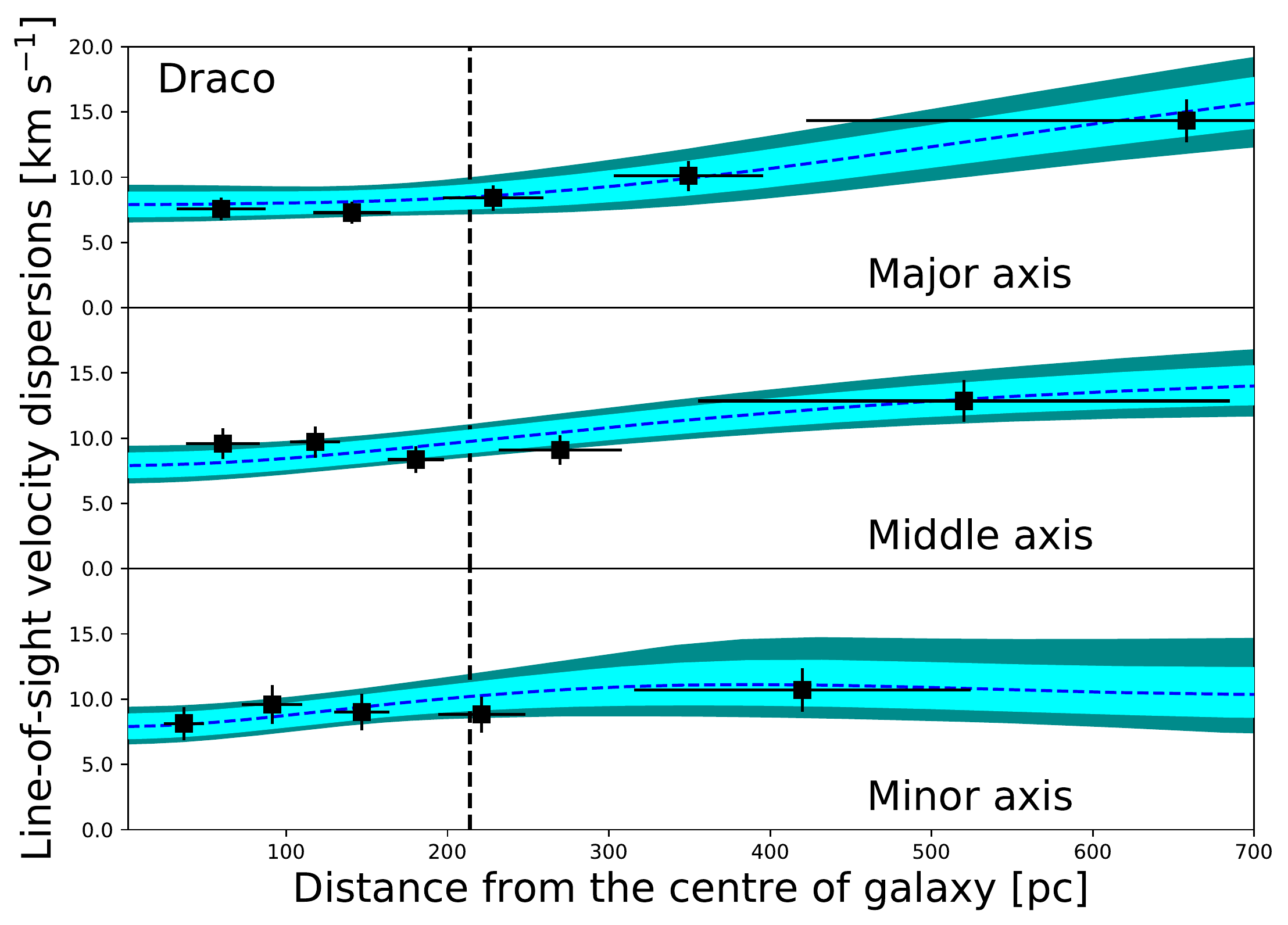}
	\includegraphics[scale=0.35]{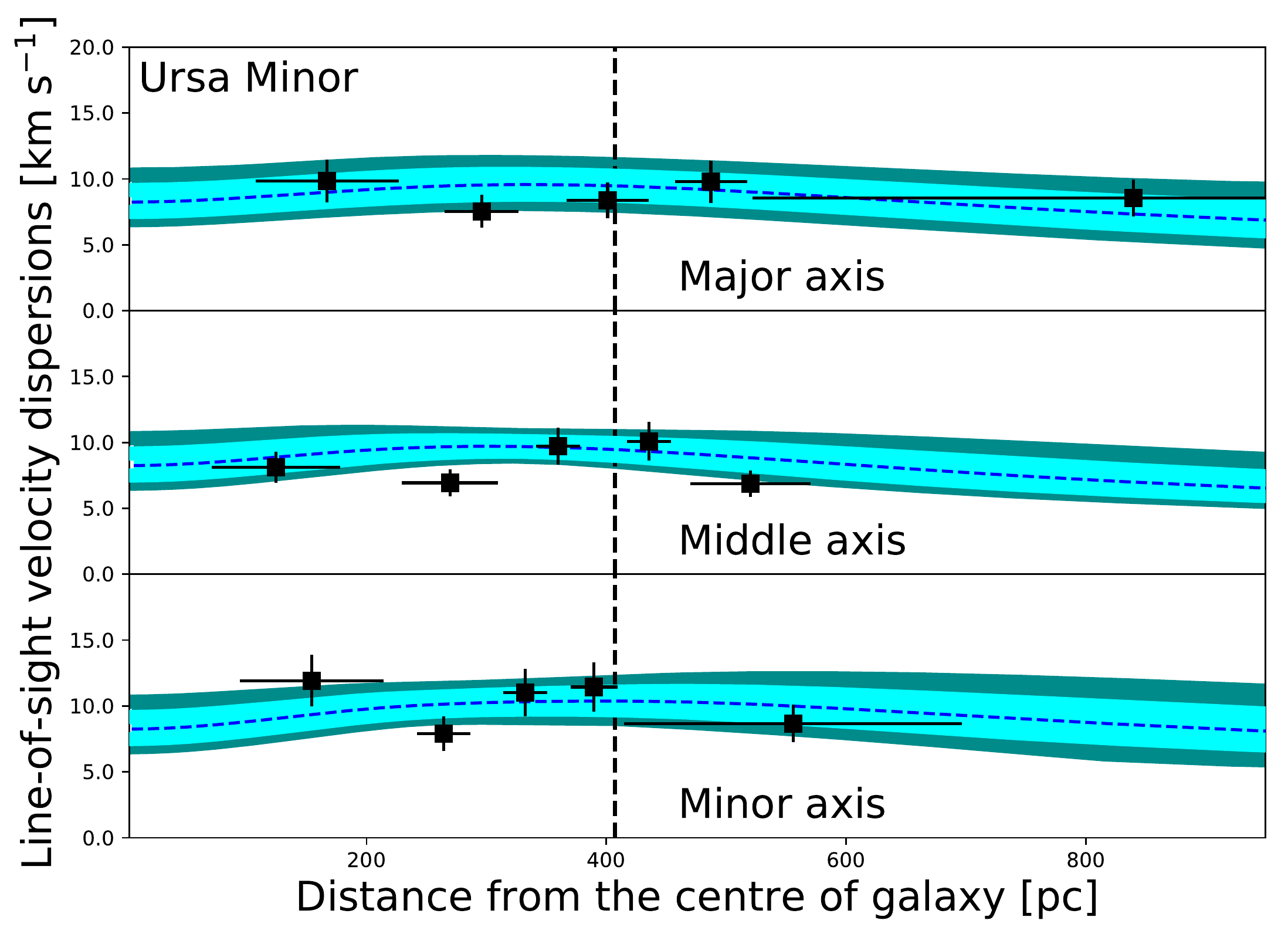}
	\includegraphics[scale=0.35]{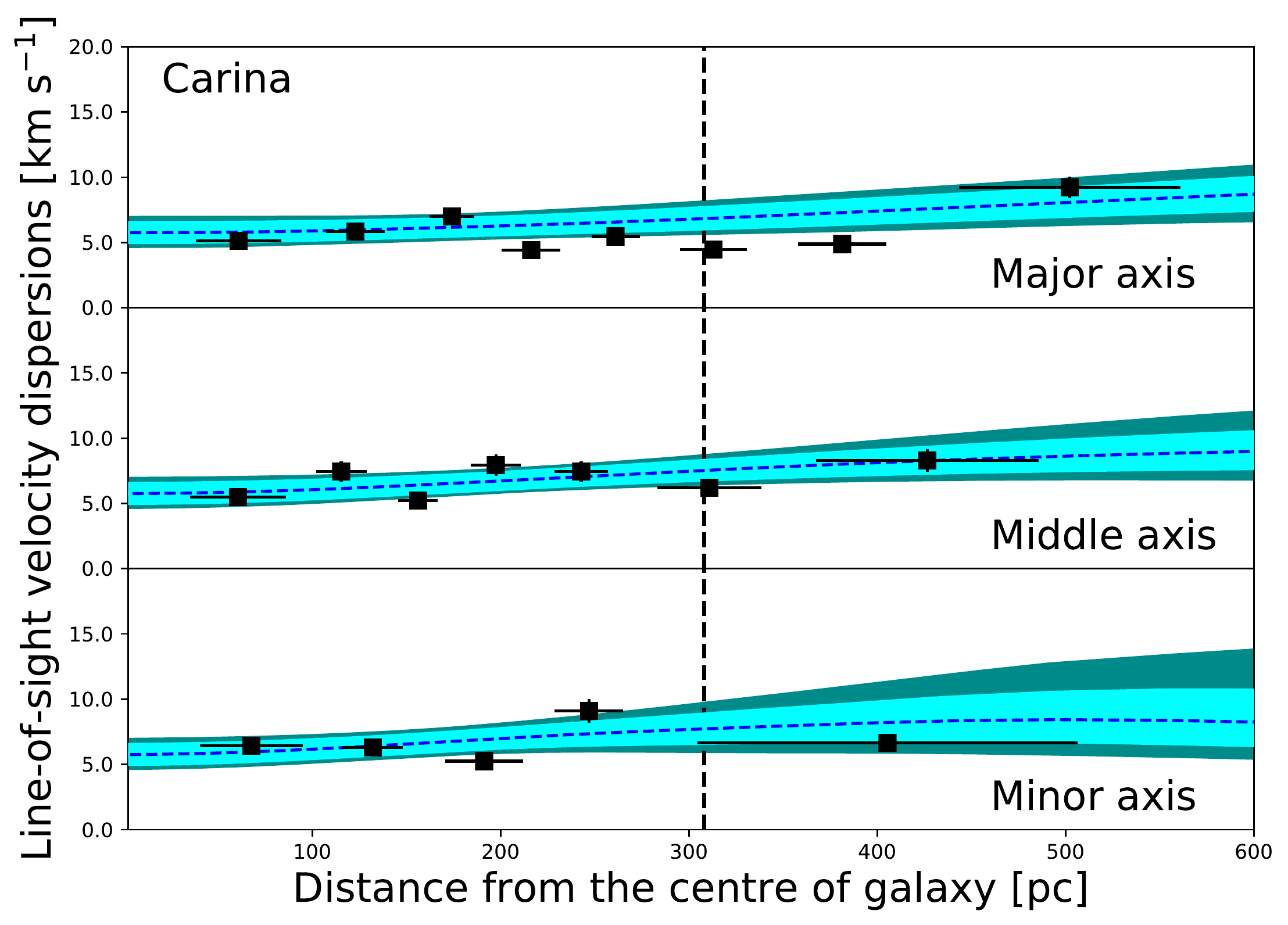}
	\includegraphics[scale=0.35]{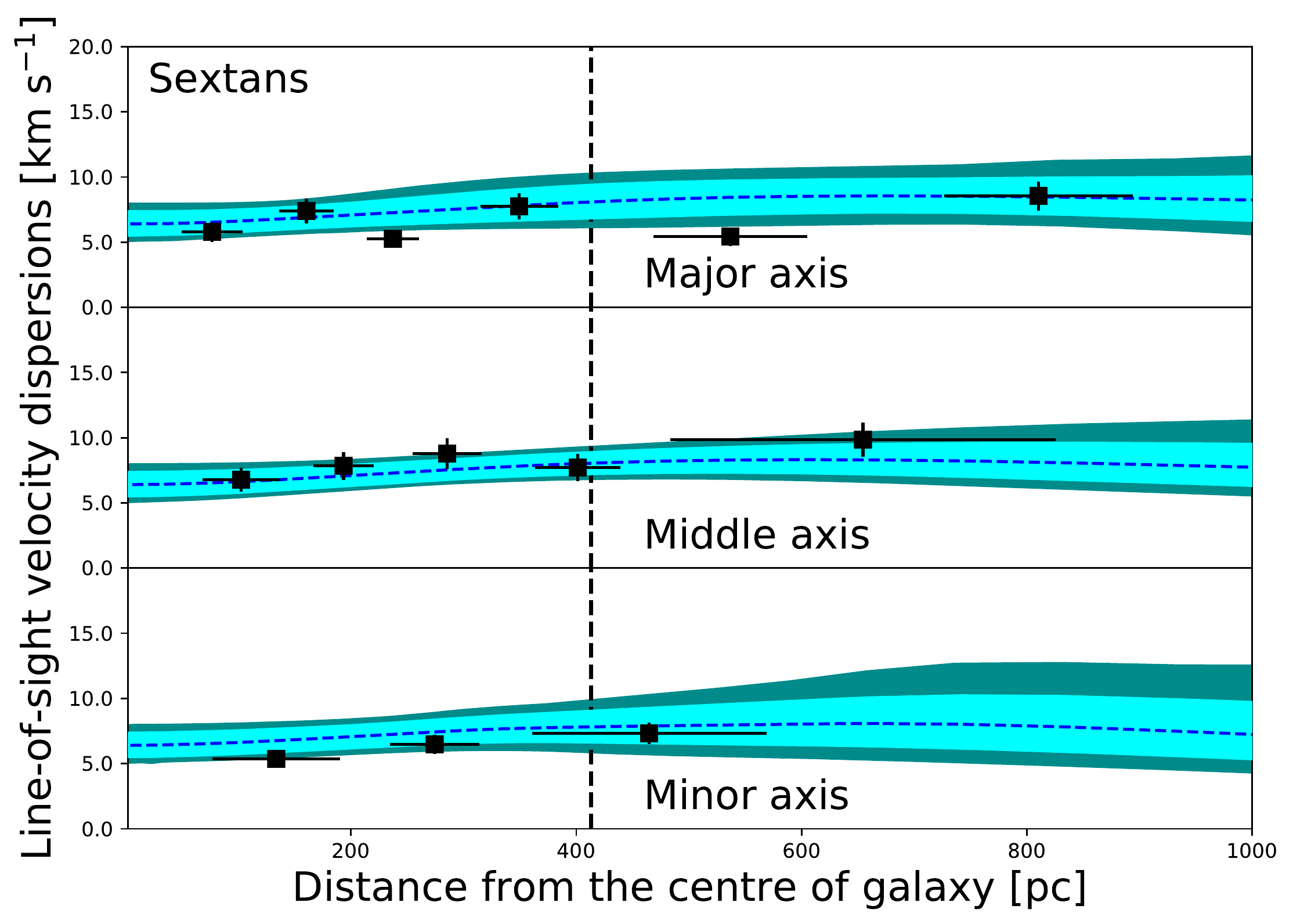}
	\includegraphics[scale=0.35]{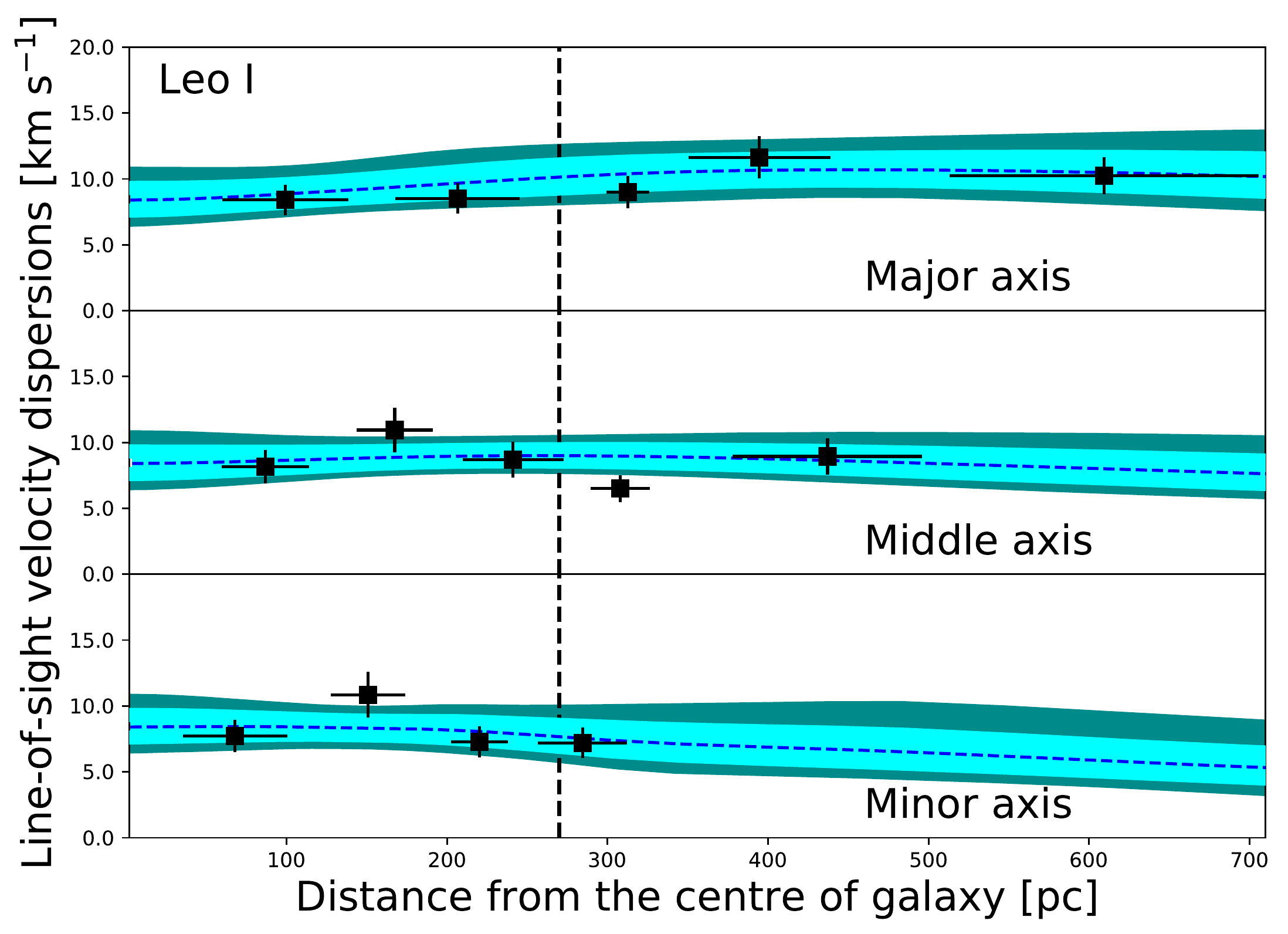}
	\includegraphics[scale=0.35]{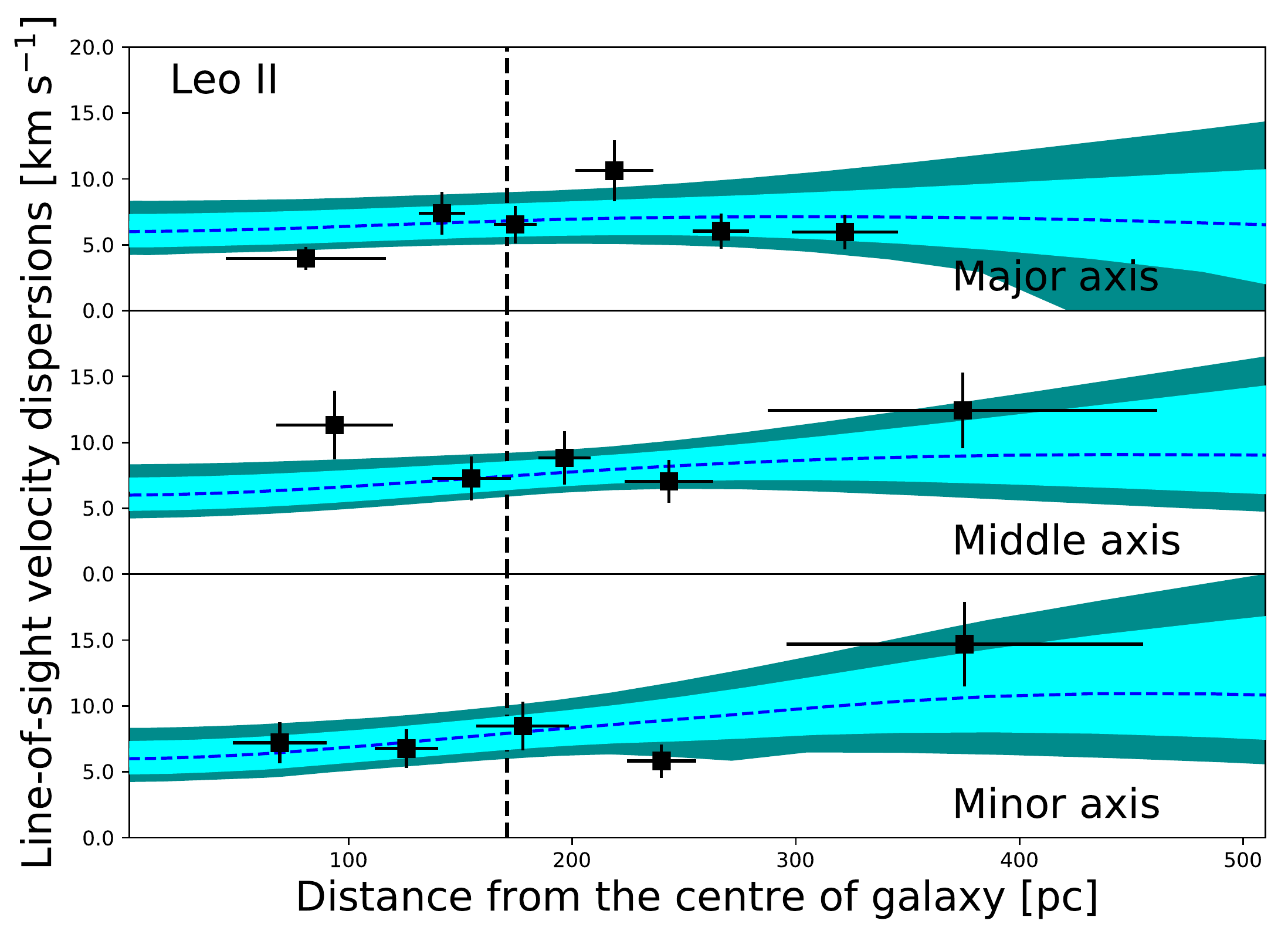}
	\includegraphics[scale=0.35]{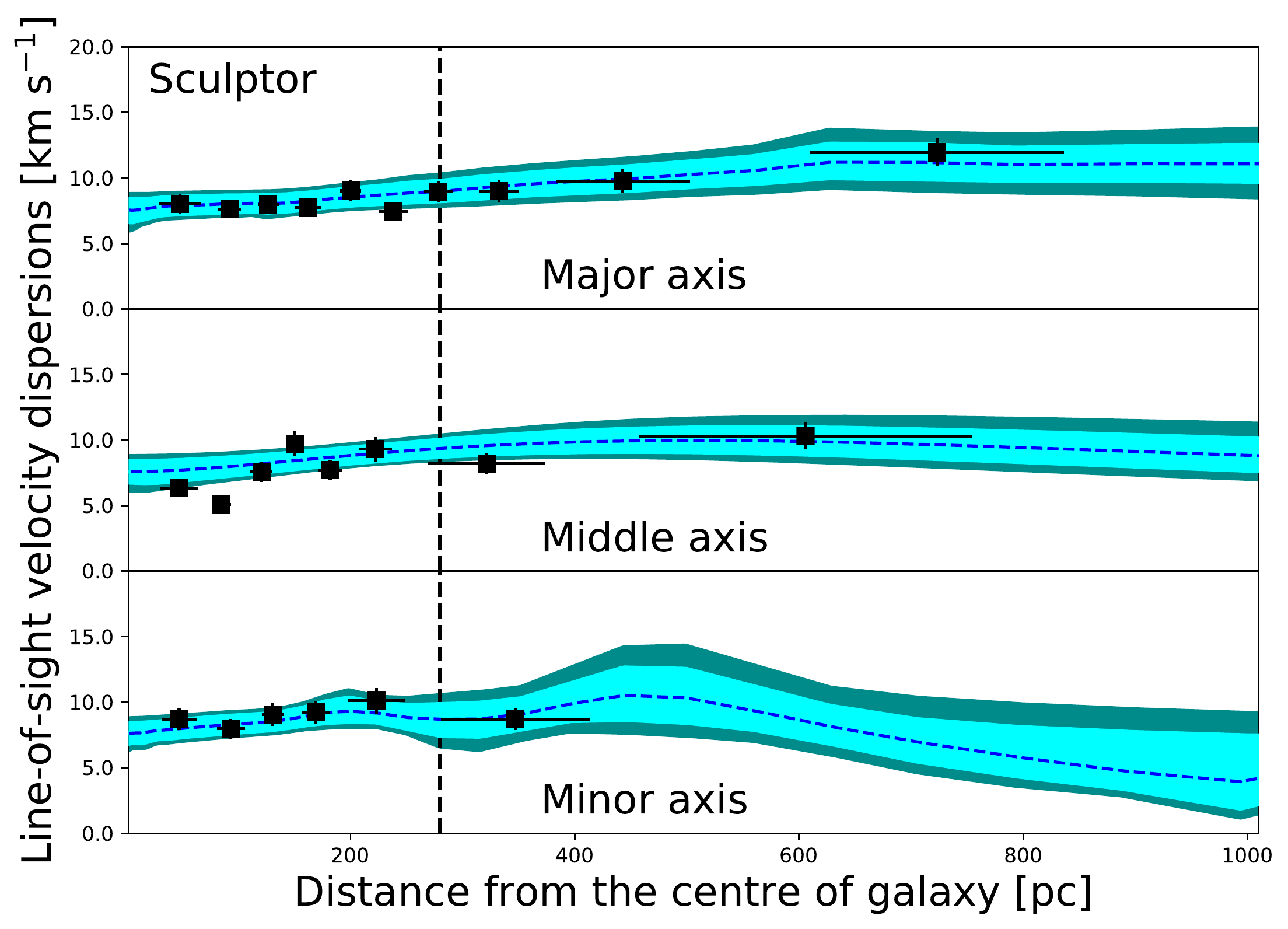}
	\includegraphics[scale=0.35]{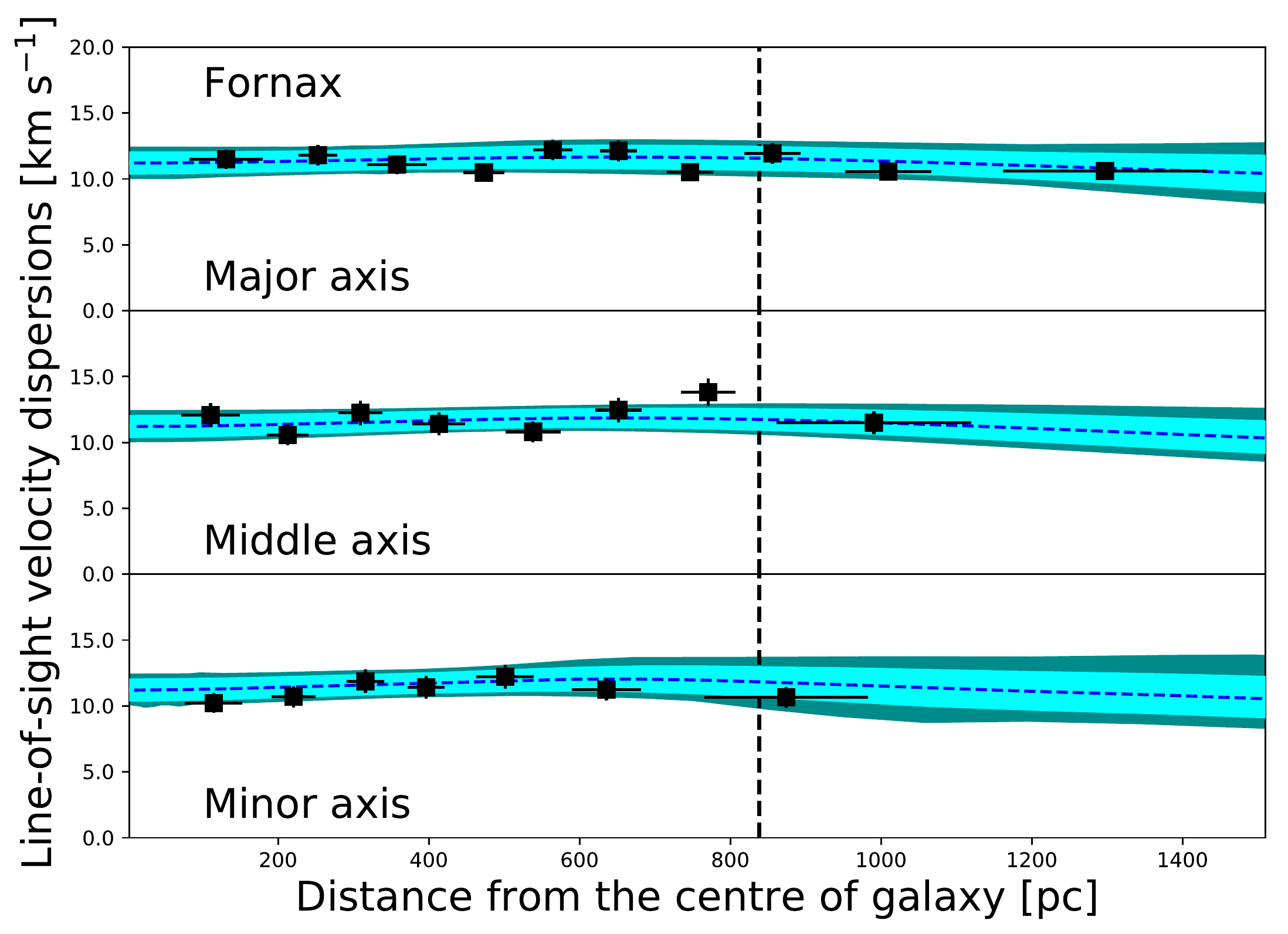}
    \caption{Line-of-sight velocity dispersion along major, middle and minor axes for each dSph. The black diamonds with error bars in each panel denote the observed ones. The dashed blue lines are the median velocity dispersion of the models and the light and dark cyan shaded regions encompass the 68 per cent and 95 per cent confidence levels from the results of the unbinned MCMC analysis. The vertical dashed lines in each panel correspond to their half-light radii.}
    \label{los}
\end{figure*}

\section{Results}
In this section, we present the results from the MCMC fitting analysis described above and some intriguing tendencies among the resultant parameters. We also show an estimation of the combined constraint by fitting all dSphs to determine a particle mass of ULA dark matter. 
\subsection{Best fit models of ULA dark matter haloes}
\subsubsection{Posterior probability distribution functions}
Figure~\ref{draumi}-\ref{sclfnx} display the samples from the posterior PDFs implemented by the fitting procedure for all dSphs.
The contours for each dSph show 68, 95 and 99.7~per~cent credible interval levels.
The vertical lines in each histogram also show median and 68~per~cent credible interval levels.
The best-fit results for each dSph are summarized in Table~\ref{table2}.
The error values correspond to the 68~per~cent credible interval.

From the posterior PDFs of all dSphs, their inclination angles $i$ are distributed in wide parameter ranges and thus it is difficult to get limits on them.
However, this uncertainty does not affect largely the constraints on the other free parameters.
Actually, our results are compatible with the results from the previous axisymmetric works~\citep{2015ApJ...810...22H,2016MNRAS.461.2914H}.

In the PDFs, there are some correlations among the parameters for all dSphs.
The most notable one is between the particle mass of ULA dark matter $m_{\psi}$ and the soliton core radius $r_c$, shown by~\citet{2017MNRAS.468.1338C}.
This degeneracy clearly arises from the relation between the soliton core density, the core radius and the ULA dark matter mass predicted by the numerical simulations~(Equation~\ref{solitoncore}).
Although this correlation appears in all dSphs, most of the dSphs except for Leo~II are well constrained. This is because the number of kinematic data of Leo~II would not be enough to place stringent constraints on them. Thus, for lack of kinematic sample volume, it is difficult to obtain tight limits on their dark matter halo parameters~\citep[see also][]{2015ApJ...810...22H}.


Another one is a degeneracy between the shape of the dark matter halo $Q$ and the stellar velocity anisotropy $\beta_z$.
This degeneracy has already been discussed in \citet{2008MNRAS.390...71C} and \citet{2015ApJ...810...22H}, who showed that the variation of these parameters has a similar impact on the central line-of-sight velocity dispersion profiles.
Besides, \citet{2015ApJ...810...22H} argued that an adequate kinematic data of stars in the outer region of a galaxy can provide a systematic difference of these parameters on the outer line-of-sight velocity dispersion profiles and thus break the degeneracy between them~(see Figure~12 in their paper).
We will explain the mechanism how this degeneracy occurs in the later part of this section.

Furthermore, there is a weak correlation between $Q$ and $r_c$ (or $m_{\psi}$), which means that smaller $Q$ has a similar effect to a smaller (larger) $r_c$ ($m_{\psi}$). 
This is explained as follows: 
a smaller $r_c$ corresponds to a higher core density, and consequently an entire line-of-sight velocity dispersion profile becomes higher.
On the other hand, the central velocity dispersion increases at small $Q$~(i.e., $Q<1$), because it yields stronger gravitational force along the $z$-direction on the basis of equation~(\ref{AGEb03}) and (\ref{AGEb04}).

\subsubsection{Line-of-sight velocity dispersion profiles}
Figure~\ref{los} shows the line-of-sight velocity dispersion profiles along the projected major, minor and middle~(which is defined at $45^{\circ}$ from the major axis) axes for visualization of the best-fit parameters for each dSph.
To obtain these binned profiles from the observed kinematic data, we implement the standard binning technique with three steps. First, we analyze the line-of-sight velocity data by folding the stellar distribution into the first quadrant in each dSph under the assumption of an axisymmetric system. 
Second, we transform the sky coordinates~$(x,y)$ (of the first quadrant) to the two-dimensional polar coordinates~$(r,\theta)$, where $\theta=0^{\circ}$ is defined along the major axis. Then we divide this into three regions in increments of $30^{\circ}$ in the direction from $\theta=0^{\circ}$ to $90^{\circ}$. Expedientially, the region $\theta=0^{\circ}$-$30^{\circ}$ is labeled as a major axis area, $\theta=30^{\circ}$-$60^{\circ}$ as a middle axis area, and $\theta=60^{\circ}$-$90^{\circ}$ as a minor axis area. Third, for each region, we radially separate stars into bins, which is comprised of a nearly equal number of stars, and then calculate the velocity dispersion with respect to each bin: $\sim100$ stars/bin for Fornax, $\sim60$ stars/bin for Carina and Sculptor, $\sim30$ stars/bin for Sextans and Draco, $\sim20$ stars/bin for Ursa~Minor, Leo I, and $\sim10$ stars/bin for Leo II.
In this figure, the blue dashed lines and shaded regions denote the median and credible interval levels~(light cyan: 68~per~cent, dark cyan: 95~per~cent) computed from the posterior PDFs of the parameters by {\it unbinned} MCMC analysis, whilst the black squares with error bars are {\it binned} velocity dispersions estimated from the observed data. These errors correspond to the 68~per~cent confidence interval.
It is found from the figure that our unbinned analysis can reproduce even binned line-of-sight velocity dispersion profiles. 

\subsubsection{Flattened shapes of dark matter haloes}
Despite the presence of several correlations among the parameters, the axial ratios of dark matter haloes in most dSphs are plausible to be elongated and oblate shapes~(i.e. $Q<1$).
In particular, Draco favours a strongly elongated dark matter halo.
Why does this galaxy tend to have an elongated dark matter halo? 
We schematically illustrate this reason in Figure~\ref{demolos}-\ref{demolos3} in the appendix. 
Given that a galaxy has a flattened stellar distribution, their $\sigma_{\rm los}$ profiles along the major and minor axes have trough- and crest-like features from the central to outer parts of the system. 
However, the $\sigma_{\rm los}$ profiles obtained from the observational data appear to be almost flat or to increase gradually toward its outskirts even in the flattened stellar distributions. 
In particular, Draco's $\sigma_{\rm los}$ profile along the major axis is flat within its half-light radius and then clearly increases toward the outer region, while that along the minor axis is almost flat profile~(see Figure~\ref{los}).

In order to reproduce such observed velocity dispersion profiles, (i)~more flattened~(i.e., $Q<1$) dark matter halo or (ii)~steeper inner slope of dark matter halo~(such as an NFW profile) or (iii)~radially-biased stellar velocity anisotropy~($\beta_z>0$) or (iv)~larger $r_c$ would be required.
Firstly, in the framework of ULA dark matter models, all of the numerical simulations predict that all dark matter haloes inevitably have cored profiles.
Therefore, the method (ii) is not able to change any velocity dispersion profiles.
Secondly, according to the top-middle panel in Figure~\ref{demolos}, a radially biased $\beta_z$ can increase the central velocity dispersion but decrease it at outer parts, simultaneously.
Thus, (iii) would be capable of reproducing the flat velocity dispersion profiles inferred by most of the dSphs. 
However, the effects of radially biased $\beta_z$ are difficult to reproduce an upward velocity dispersion profile along the major axis toward its outer part like Draco.
Although larger $r_c$ can help to reproduce such an upward profile~(see the upper-middle panels in Figure~\ref{demolos2} and \ref{demolos3}), this effect on a $\sigma_{\rm los}$ profile is inconsistent with the observed flat profile along the minor axis~(see the lower-middle panels in Figure~\ref{demolos2} and \ref{demolos3}).
Thirdly, from the upper-left panels in Figure\ref{demolos}-\ref{demolos3}, the effects of decreasing $Q$ from unity to $0.5$ can roughly reproduce a flat or an upward dispersion profile. Meanwhile, when a dark matter halo becomes more elongated such as $Q<0.5$, the features of $\sigma_{\rm los}$ profiles change drastically, and these profiles are quite different from the observed ones.
However, in combination with the effects of a tangentially-biased velocity anisotropy~($\beta_z<0$) and a large core radius of ULA dark matter halo~($r_c/b_{\ast}>1$), the shapes of $\sigma_{\rm los}$ profiles along the major and minor axes are in good agreement with the observed $\sigma_{\rm los}$ profiles even with a very elongated dark matter halo~(see light and dark orange lines in the right panels of Figure~\ref{demolos2} and \ref{demolos3}).
Furthermore, the parameters which we draw the dark orange lines in the right panels of Figure~\ref{demolos3} are roughly consistent with the best-fit values of $Q$, $\beta_z$ and $r_c/b_{\ast}$ for Draco~(Table~\ref{table2}).
Consequently, it is found that a flattened ULA dark matter halo and a combination of the three parameters can be of very importance in reproducing the observed flat or upward dispersion profiles along the major and minor axes.

\begin{figure*}
	\includegraphics[scale=0.7]{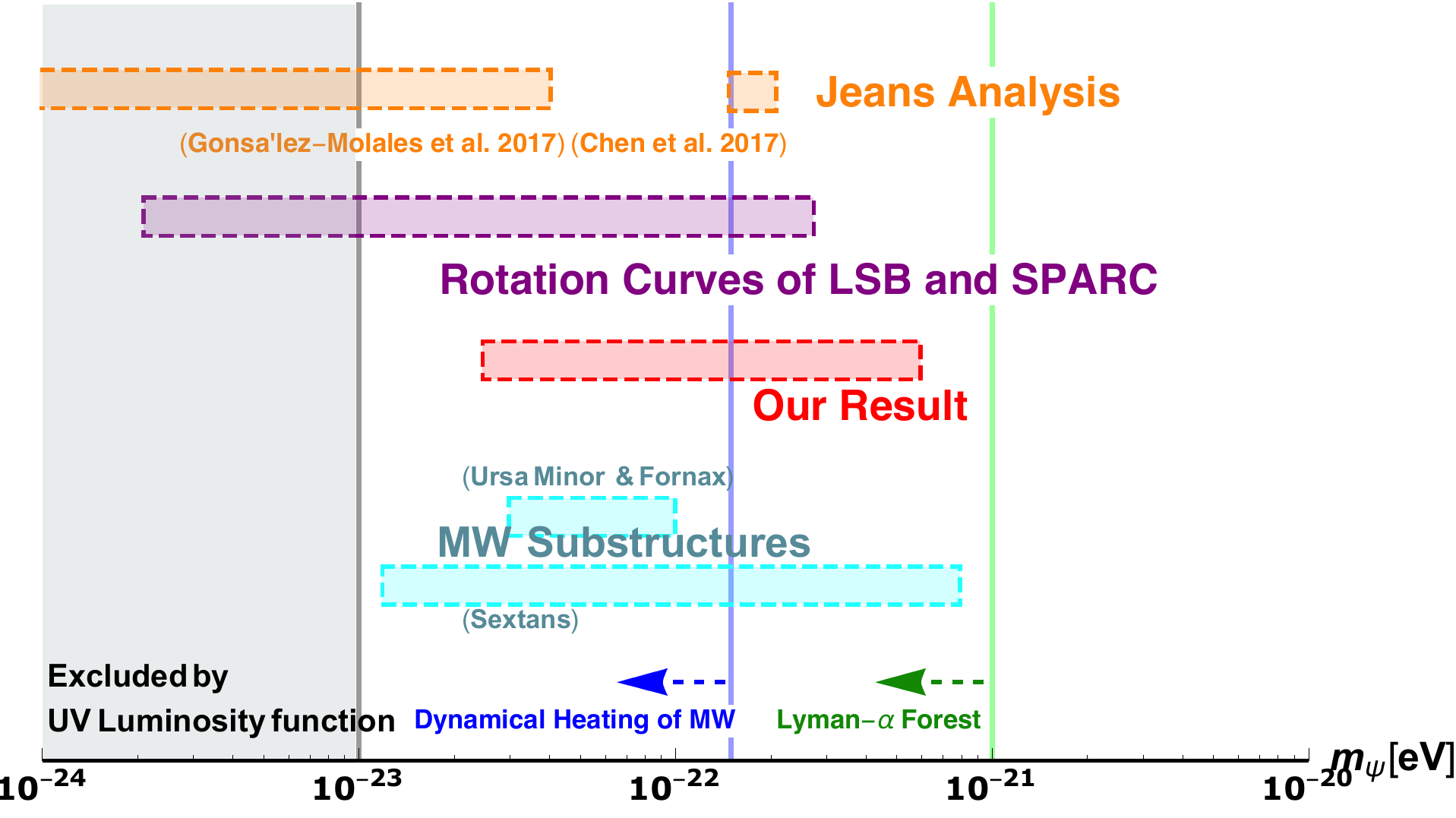}
    \caption{Constraints on the ULA mass from our study ($m_{\psi}=1.05^{+4.98}_{-0.80}\times10^{-22}$~eV, 2$\sigma$ level) with red contour compared with the other mass bands favoured by several measurements such as previous Jeans analysis works (\citet{2017MNRAS.472.1346G}, \citet{2017MNRAS.468.1338C}, orange), presences of MW substructures (\citet{2012JCAP...02..011L}, \citet{2014JCAP...09..011L}, cyan), rotation curves of LSB and SPARC galaxies (\citet{2018MNRAS.475.1447B}, purple).
    The lower mass regions less than blue and green line are disfavoured by dynamical heating of stellar streams in MW (\citet{2018arXiv180800464A}) and Lyman-$\alpha$ forest ($m_{\psi} \gtrsim 10^{-21}$ eV, green) at around 2$\sigma$ levels. The lower mass region $m_\psi \lesssim 10^{-23}$ eV is severely excluded by the UV Luminosity function of high-$z$ galaxies at more than 8$\sigma$ level \citep{2015MNRAS.450..209B}.}
    \label{new}
\end{figure*}

\subsection{The combined constraint on particle mass of ULA dark matter}
One of our purposes in this work is to obtain a limit on the mass of ULA dark matter particle.
Since the value of $m_{\psi}$ would be universal in the Universe, we can perform the joint analysis by fitting all dSphs simultaneously.
Following a hierarchical model introduced by~\citet{2015MNRAS.451.2524M}, we suppose that each kinematic data of dSphs are statistically independent, and hence the joint likelihood function is written by the production of the likelihood functions of the dSphs~(Equation~\ref{LF}) described as,
\begin{equation}
{\cal L}_{\text{joint}} = \prod^{N_{\text{dSph}}}_{k=1}\prod^{N^{k}_{\text{data}}}_{j=1}{\cal L}^{k}_{j}.
\end{equation}
While $m_{\psi}$ is regarded as a common parameter, $Q$, $r_c$, $\beta_z$ and $i$ are allowed to vary in individual dSphs~\citep[see also][but using spherical mass model]{2017MNRAS.468.1338C}.
As a result, we obtain $1\sigma$ $(2\sigma)$ confidence intervals of $\log_{10}([m_{\psi}/10^{-23}{\text{eV}}])=0.94^{+0.51 (+0.75)}_{-0.42 (-0.61)}$.
In comparison with the results of the individual dSphs~(Table~\ref{table2}), this suggests that the constraint on $m_{\psi}$ would depend largely on the results from Fornax and Sculptor which have sufficiently large number of data sample.

\begin{table*}
	\centering
	\caption{Parameter constraints for MW dSph satellites. The dark halo potential is assumed to be soliton+NFW profile. Errors correspond to the $1\sigma$ range of our analysis. $r_{\epsilon}$ is a transition radius from soliton to NFW profile calculated from Equation~(\ref{Rtran}). $\rho_c$ and $\rho_s$ are the central densities of soliton core and NFW profiles, respectively.}
	\label{table3}
	\scalebox{0.92}[1.1]{
	\begin{tabular}{ccccccccccc} 
		\hline\hline
Object   & $Q$ & $\log_{10}(r_{c})$ & $\log_{10}$($m_{\psi})$ & $-\log_{10}(1-\beta_z)$ & $\log_{10}(\epsilon)$  & $\log_{10}(r_s)$ &  $i$ & $\log_{10}(r_{\epsilon})$ & $\log_{10}(\rho_c)$& $\log_{10}(\rho_s)$\\
&  & [pc] & [$10^{-23}{\text eV}$] &  & & [pc] & [deg] & [pc] & [$M_{\odot}$~kpc$^{-3}$]& [$M_{\odot}$~kpc$^{-3}$]\\
		\hline
Draco       & $ 0.29 ^{+ 1.12 } _{- 0.11 } $ & $ 2.75 ^{+ 0.19 } _{- 1.20 } $ & $ 0.84 ^{+ 1.82 } _{- 0.37 } $ & $ -0.09 ^{+ 0.49 } _{- 0.15 } $ & $ -1.25 ^{+ 0.55 } _{- 1.12 } $ & $ 2.62 ^{+ 1.19 } _{- 1.1 } $ & $ 71.47 ^{+ 12.32 } _{- 13.4 } $ & $ 3.06 ^{+ 0.39 } _{- 1.12 }$ & $ 8.65 ^{+ 1.04 } _{- 0.11 } $ & $ 9.07 ^{+ 2.93 } _{- 2.18 } $\\
Ursa~Minor  & $ 1.41 ^{+ 0.39 } _{- 0.55 } $ & $ 1.52 ^{+ 0.59 } _{- 0.60 } $ & $ 2.50 ^{+ 0.32 } _{- 0.49 } $ & $ 0.62 ^{+ 0.17 } _{- 0.13 } $  & $ -1.20 ^{+ 0.84 } _{- 2.41 } $ & $ 2.73 ^{+ 0.35 } _{- 0.32 } $& $ 78.88 ^{+ 7.12 } _{- 7.34 } $  & $ 1.77 ^{+ 0.42 } _{- 0.18 } $& $ 9.78 ^{+ 2.89 } _{- 0.82 } $ & $ 7.88 ^{+ 0.57 } _{- 0.35 } $\\
Carina      & $ 0.45 ^{+ 0.37 } _{- 0.20 } $ & $ 2.69 ^{+ 0.18 } _{- 0.78 } $ & $ 1.23 ^{+ 1.22 } _{- 0.37 } $ & $ 0.17 ^{+ 0.21 } _{- 0.18 } $  & $ -1.10 ^{+ 0.50 } _{- 0.49 } $ & $ 2.11 ^{+ 1.54 } _{- 0.80 } $& $ 71.13 ^{+ 12.31 } _{- 11.92 } $& $ 3.01 ^{+ 0.02 } _{- 0.15 } $& $ 8.10 ^{+ 0.65 } _{- 0.20 } $ & $ 10.02 ^{+ 2.06 } _{- 3.36 } $\\
Sextans     & $ 0.70 ^{+ 0.66 } _{- 0.35 } $& $ 2.73 ^{+ 0.17 } _{- 0.71 } $& $ 1.25 ^{+ 0.98 } _{- 0.34 } $& $ 0.15 ^{+ 0.20 } _{- 0.17 } $& $ -2.26 ^{+ 1.59 } _{- 1.81 } $& $ 2.82 ^{+ 0.79 } _{- 1.10 } $& $ 68.66 ^{+ 13.68 } _{- 10.78 } $& $ 3.31 ^{+ 0.17 } _{- 0.71 } $& $ 7.94 ^{+ 0.75 } _{- 0.16 } $& $ 7.39 ^{+ 2.67 } _{- 0.87 } $\\
Leo~I       & $ 0.81 ^{+ 0.72 } _{- 0.47 } $& $ 1.77 ^{+ 0.63 } _{- 0.74 } $& $ 2.33 ^{+ 0.48 } _{- 0.84 } $& $ 0.08 ^{+ 0.23 } _{- 0.18 } $& $ -1.08 ^{+ 0.66 } _{- 2.34 } $& $ 2.71 ^{+ 0.55 } _{- 0.48 } $& $ 67.3 ^{+ 14.63 } _{- 12.55 } $& $ 1.91 ^{+ 0.72 } _{- 0.32 } $& $ 9.42 ^{+ 2.30 } _{- 0.58 } $& $ 8.14 ^{+ 0.80 } _{- 0.55 } $\\
Leo~II      & $ 1.13 ^{+ 0.56 } _{- 0.63 } $& $ 2.38 ^{+ 0.40 } _{- 1.07 } $& $ 1.68 ^{+ 0.95 } _{- 0.69 } $& $ 0.17 ^{+ 0.19 } _{- 0.25 } $& $ -2.62 ^{+ 1.78 } _{- 1.62 } $& $ 2.93 ^{+ 0.69 } _{- 1.14 } $& $ 57.02 ^{+ 21.21 } _{- 15.7 } $& $ 2.95 ^{+ 0.39 } _{- 1.10 } $& $ 8.45 ^{+ 2.20 } _{- 0.21 } $&$ 7.33 ^{+ 2.43 } _{- 1.00 } $\\
Sculptor    & $ 0.36 ^{+ 0.26 } _{- 0.13 } $& $ 2.64 ^{+ 0.1 } _{- 0.11 } $& $ 1.02 ^{+ 0.29 } _{- 0.17 } $& $ 0.22 ^{+ 0.22 } _{- 0.23 } $& $ -1.48 ^{+ 0.49 } _{- 0.47 } $& $ 2.13 ^{+ 1.14 } _{- 0.68 } $& $ 55.13 ^{+ 22.1 } _{- 3.11 } $& $ 3.02 ^{+ 0.12 } _{- 0.11 } $& $ 8.70 ^{+ 0.17 } _{- 0.29 } $&$ 10.03 ^{+ 1.60 } _{- 1.87 } $\\
Fornax      & $ 0.53 ^{+ 0.51 } _{- 0.26 } $& $ 1.40 ^{+ 1.55 } _{- 0.34 } $& $ 2.69 ^{+ 0.23 } _{- 1.85 } $& $ 0.11 ^{+ 0.15 } _{- 0.14 } $& $ -1.94 ^{+ 1.2 } _{- 0.9 } $& $ 3.59 ^{+ 0.27 } _{- 1.06 } $& $ 68.61 ^{+ 12.49 } _{- 10.59 } $& $ 1.72 ^{+ 1.67 } _{- 0.11 } $& $ 9.94 ^{+ 1.65 } _{- 1.88 } $&$ 6.93 ^{+ 1.77 } _{- 0.21 } $\\
	\hline
	\end{tabular}
	}
\end{table*}

\section{Discussion}
\subsection{Comparison with other works}
In this section, we compare our results with the other studies based on the Jeans analysis as well as other astronomical phenomena.
Figure~\ref{new} shows the constraints on the ULA dark matter mass from this work in comparison with those from the other previous works. In what follows, we will describe these limits in details.

Following to the methods developed by~\citet[][hereafter WP11]{2011ApJ...742...20W}, S14~utilized the multiple chemo-dynamical stellar populations of Fornax and a spherical Jeans analysis to determine $m_{\psi}$, and then they obtained $m_{\psi}=8.1^{+1.6}_{-1.7}\times10^{-23}$~eV.
\citet{2015MNRAS.451.2479M} also performed the similar analysis for multiple stellar populations of Sculptor as well as Fornax.
They concluded that in order for ULA dark matter to resolve core/cusp problem, the particle mass should be less massive than the upper limit, $m_{\psi}<1.1\times10^{-22}$~eV at 95 per~cent confidence level.
Similarly, \citet{2017MNRAS.472.1346G} made an attempt to place constraints on the upper limit of the particle mass through the WP11's method.
Unlike the above works, however, they adopted not the mean stellar velocity dispersion of the different stellar populations but the luminosity-averaged ones, because these could provide unbiased constraints on the dark matter halo parameters.
Then, their unbiased analysis led to a severer limit to the ULA dark matter mass, $m_{\psi}<0.4\times10^{-22}$~eV at 97.5~per~cent confidence.
This limit is in tension with the other constraints such as the Jeans analysis and a subhalo mass function in MW-like galaxies.
However, as shown recently by \citet{2018MNRAS.474.1398G}, using dSph-like galaxies from APOSTLE simulation, the WP11's method is even sensitive to the viewing angle used in the kinematic sample. In particular, if a dSph is not spherical, the assumption of spherical symmetry in this method may lead to a strong bias. Therefore, we should bear in mind that this method might introduce large uncertainties on the estimate of a dark matter density profile.
On the other hand, \citet{2017MNRAS.468.1338C} applied the full spherical Jeans analysis assuming a soliton core dark matter profile to the line-of-sight velocity dispersion profiles calculated from the current available kinematic data of eight luminous dSphs and obtained $m_{\psi}=1.79^{+0.35}_{-0.33}\times10^{-22}$~eV at approximately 95~per~cent confidence.
Furthermore, they also estimated independently these parameters for Sculptor and Fornax through their multiple stellar subpopulations.
Then, they obtained $\log_{10}([m_{\psi}/10^{-23}{\text{eV}}])=1.08^{+0.28}_{-0.22}$~(assummed by Osipkov-Merritt velocity anisotropy) for Sculptor and  $\log_{10}([m_{\psi}/10^{-23}{\text{eV}}])=0.91^{+0.26}_{-0.09}$ for Fornax, respectively. Actually, our estimated ULA dark matter masses for these galaxies are consistent with their constraints.

Comparing our best-fitting particle mass~($m_{\psi}=0.87^{+4.03}_{-0.66}\times10^{-22}$~eV at 95~per~cent confidence) with the above previous works, our inferred mass limits (especially the upper limit) are less stringent than those by the previous spherical works.
The main reason for this difference is that our axisymmetric analysis fully takes into account the uncertainties of the dark matter halo shape and the inclination angle.
Additionally, as mentioned above, the WP11's method imposes that the stellar and dark components are spherical symmetry. If these components in a dSph are not spherical, then the spherical model may lead to a strong bias and the inferred slope turns out to depend largely on the line of sight~\citep{2013MNRAS.431.2796K,2018MNRAS.474.1398G}.
Therefore, this method still has large systematic uncertainties in the constraints on ULA dark matter mass $m_{\psi}$. 

Low surface brightness~(LSB) galaxies are also suitable objects to obtain limits on density profiles of dark matter haloes.
This is because these galaxies are the dark matter dominated systems similar to the Galactic dSphs, and there exist their accurate rotation curve data~ \citep[e.g.,][]{2016AJ....152..157L}.
\citet{2018MNRAS.475.1447B} made an attempt to reproduce these rotation curves of high-resolution LSB galaxies using ULA dark matter models. From the fitting results with assuming a soliton~$+$~NFW dark matter profile, they obtained $0.212 < m_{\psi}/(10^{-23}{\rm eV}/c^2)<27.0$ with soliton core radius~$0.326<r_c/{\rm kpc} < 8.96$. This constraint is overlapped with our result (Figure~\ref{new}).

Next, we compare with the constraints from other individual observables on various spacial scales from ultra-faint dwarf galaxies to CMB.
First, the precision CMB data, which relies only on linear scale regime, can place constraints on ULA dark matter mass.
\citet{2018MNRAS.476.3063H} found that the temperature anisotropies, E-mode polarization and lensing deflection derived from the full {\it Planck} data set require $m_{\psi}\gtrsim10^{-24}$~eV.
Second, \citet{2015MNRAS.450..209B} developed the luminosity functions of high redshift galaxies estimated from the deep photometric data in {\it Hubble Ultra Deep Field} to predict the reionization history of the Universe.
Comparing with the luminosity functions predicted by ULA dark matter, which can suppress the structure formation below Jeans mass scales of ULA, they ruled out $m_{\psi}\lesssim10^{-23}$~eV at more than $8\sigma$ significance~\citep[see also,][]{2016JCAP...04..012S,2016ApJ...818...89S,2017PhRvD..95h3512C}. 
Third, the flux power spectrum of Lyman-$\alpha$ forest data is also a strong tool for constraining on ULA dark matter mass.
Recently, several studies with this method obtained strong limits on $m_{\psi}\gtrsim10^{-21}$~eV~\citep[e.g.,][]{2017MNRAS.471.4606A,2017PhRvL.119c1302I,2017PhRvD..96l3514K,2019MNRAS.482.3227N}.
Taken at face value, our results are roughly consistent with these constraints, except for the Lyman-$\alpha$ forest.
We should, however, keep in mind that Lyman-$\alpha$ constraints are highly dependent upon the modelling for baryonic effects and data~\citep{2017PhLB..773..258G,2017PhRvD..95d3541H}.

Besides, the MW substructures such as stellar streams and ultra-faint dwarf galaxies also enable us to constrain a particle mass of ULA as dark matter.
For instance, the existence of cold clumps and globular clusters in the MW dSphs require $m_{\psi}\sim0.3-1.0\times10^{-22}$~eV for Ursa Minor and Fornax dSphs~\citep{2012JCAP...02..011L} and
$m_{\psi}\sim0.12-8.0\times10^{-22}$~eV for Sextans~\citep{2014JCAP...09..011L}.
Also, based on the thickening of the MW disk caused by dynamical heating from ULA dark matter substructures, \citet{2018arXiv180904744C} provided a lower limit on ULA mass, $m_{\psi}\gtrsim0.6\times10^{-22}$~eV.
Similarly, using the thickening of the MW stellar streams stemmed from quantum fluctuations of density field of ULA dark matter haloes, \citet{2018arXiv180800464A} obtained a conservative lower limit $m_{\psi}>1.5\times10^{-22}$~eV.
More recently, \citet{2018arXiv181008543M} indicated that using the existence of a old star cluster within the central region of Eridanus~II which is a newly discovered ultra-faint dwarf galaxy, the upper limit of ULA mass can be inferred as $m_{\psi}\lesssim10^{-19}$~eV.
In comparison with these works on non-linear regimes, our result is not incompatible with their constraints (see Figure~\ref{new}).

\subsection{Soliton~$+$~NFW}

As described in Section~2.2, to justify the assumption that all stars are located within the central soliton core, we investigate the case where the soliton core connects to an NFW halo at a larger radius with respect to all of the dSphs.
To this end, we compute our non-spherical Jeans analysis with a soliton~$+$~NFW dark matter halo model written as~Equation~(\ref{solNFW}) and implement a MCMC fitting procedure described in Section 2.4.
According to Equation~(\ref{solNFW}), this density model introduces two additional parameters $(\epsilon,r_s)$, which we model with Jeffreys priors over the ranges~$-5\leq\log_{10}\epsilon<0$ and $1\leq\log_{10}r_{s}\leq4$.

The best-fit parameters for all dSphs in this analysis are tabulated in Table~\ref{table3}.
We also estimate the transition radius $r_\epsilon$ by substituting the obtained $r_c$ and $\epsilon$ into Equation~(\ref{Rtran}), and the central densities of the soliton core~$\rho_{c}$ and the NFW profile~$\rho_{s}$~(see the last three columns in Table~\ref{table3}).
Figure~\ref{DMpro} in Appendix~B shows the dark matter density profiles estimated by their best-fit parameters.
Furthermore, we calculate the ratios of $r_{\epsilon}/b_{\ast}$ and $r_{\epsilon}/r_{90}$, and these results are tabulated in Table~\ref{table4}.
From these results, most of the dSphs~(Draco, Carina, Sextans, Leo~II and Sculptor) have larger $r_\epsilon$ than $b_{\ast}$ and $r_{90}$, and the five parameters~$(Q,r_c,m_{\psi},\beta_z,i)$ for these galaxies are roughly consistent with these estimated by the soliton-only models.

\begin{table}
	\centering
	\caption{The ratio of transition radius to half-light radius~($r_{\varepsilon}/b_{\ast}$) and the ratio of transition radius to the radius that encloses 90~per~cent of the spectroscopic sample~($r_{\varepsilon}/r_{90}$).}
	\label{table4}
	\scalebox{1.2}[1.2]{
	\begin{tabular}{ccc} 
		\hline\hline
Object            & $r_{\varepsilon}/b_{\ast}$  &  $r_{\varepsilon}/r_{90}$\\
		\hline
Draco             & $ 5.34 ^{+ 7.87 } _{- 4.94 } $      & $ 2.17 ^{+ 3.20 } _{- 2.01 } $\\
Ursa~Minor        & $ 0.15 ^{+ 0.24 } _{- 0.05 } $      & $ 0.11 ^{+ 0.17 } _{- 0.04 } $\\
Carina            & $ 3.30 ^{+ 0.14 } _{- 0.97 } $      & $ 2.37 ^{+ 0.10 } _{- 0.70 } $\\
Sextans           & $ 4.99 ^{+ 2.46 } _{- 4.02 } $      & $ 2.88 ^{+ 1.42 } _{- 2.32 } $\\
Leo~I             & $ 0.30 ^{+ 1.30 } _{- 0.16 } $      & $ 0.16 ^{+ 0.71 } _{- 0.09 } $\\
Leo~II            & $ 5.26 ^{+ 7.72 } _{- 4.84 } $      & $ 2.56 ^{+ 3.76 } _{- 2.36 } $\\
Sculptor          & $ 3.30 ^{+ 0.14 } _{- 0.97 } $      & $ 2.37 ^{+ 0.10 } _{- 0.70 } $\\
Fornax            & $ 0.06 ^{+ 2.92 } _{- 0.01 } $      & $ 0.05 ^{+ 2.49 } _{- 0.01 } $\\
	\hline
	\end{tabular}
	}
\end{table}

\begin{figure*}
	\begin{center}
		\includegraphics[scale=0.45]{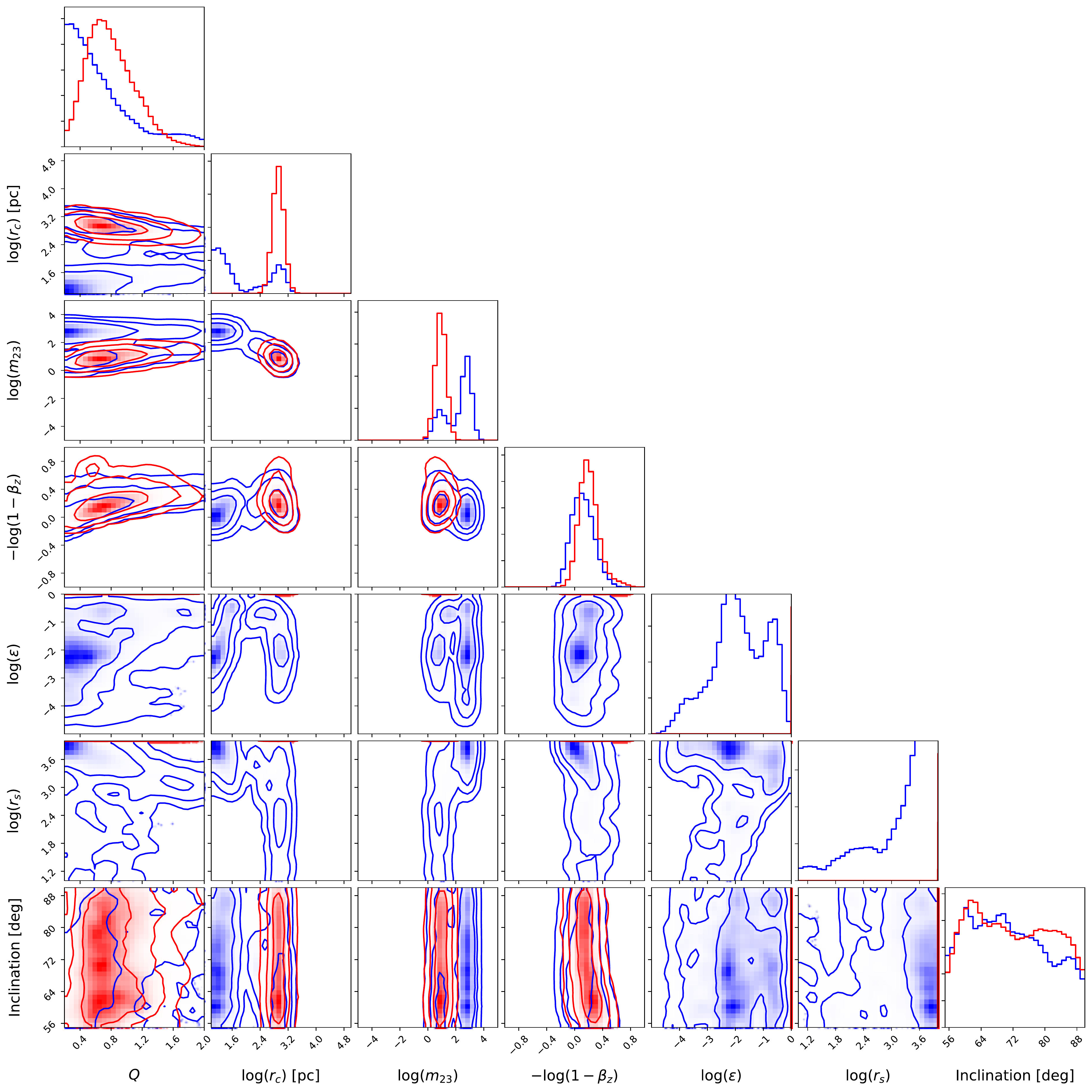}
	\end{center}
   \caption{Posterior distribution functions of dark matter halo parameters for Fornax in the case for soliton only~(red) and for soliton + NFW~(blue). $m_{23}$ is particle mass of ULA dark matter $m_{\psi}$ normalized by $10^{-23}$~eV, that is, $m_{23}=m_{\psi}/10^{-23}$~eV.}
    \label{SolitonNFW}
\end{figure*}

By contrast, however, it is found that for the case of Ursa~Minor, Leo~I and Fornax dSph, their transition radii are much smaller than their $b_{\ast}$ and $r_{90}$. Accompanied by this, their resultant parameters modeled by soliton~$+$~NFW are quite different from those modeled by the soliton-only model, especially $r_c$ and $m_{\psi}$.
In order to interpret the discrepancies, we compare the posterior PDFs obtained by soliton only and by soliton~$+$~NFW models.
Figure~\ref{SolitonNFW} shows the posterior PDFs of the free parameters computed by soliton only~(red) and by soliton~$+$~NFW~(blue) in the case for Fornax dSph.
From this figure, we find the clear bimodal posterior distributions of $r_c$ and $m_{\psi}$ in the case of the soliton~$+$~NFW model.
It is also found that while one of the bimodality coincides with the PDFs from the soliton-only model, another one is thought to be newly allowed realms of parameter spaces where the PDF of $r_c$ tends to be quite small due to an outer NFW dark matter profile.
Moreover, the scale radius $r_s$ prefers to be larger than the soliton core radius $r_c$, which gives us an impression that Fornax dSph could favour a small core~$+$~NFW profile with a large scale radius, even though these are quite huge uncertainties on the free parameters\footnote{We confirm that the other two dSphs also have a result similar to that of Fornax.}.
\citet{2017MNRAS.468.1338C} also investigated the case where the soliton core connects to an NFW halo, using spherical Jeans analysis. 
They concluded that the free parameters of Fornax estimated by soliton~$+$~NFW model are consistent with the soliton-only model. 
However, they impose $r_c$ on the lower limit which is $\sim100$~pc, while our analysis allows for a much smaller soliton core radius.
This treatment could be the origin of discrepancy.
Regarding the above results, note that cusp-core problem of the dSphs is still under debate~\citep[e.g.,][]{2010MNRAS.408.2364S,2013A&A...558A..35B,2019MNRAS.484.1401R}.
Therefore, in order to solve this issue, a large number of kinematic data of not only line-of-sight velocities but proper motions of stars are also needed.
In conclusion, to obtain reliable limits on a dark matter halo predicted from ULA, we emphasize that it might be necessary the dSphs to take into account an external NFW dark matter profile encompassing a soliton core.

\subsection{Flattened ULA dark matter halo}

In this work, to obtain plausible limits on ULA dark matter particle mass, we incorporate non-sphericity of the dark matter halo into dynamical mass models for the Galactic dSphs. Then it is found from our analysis that the dSphs would prefer to have a flattened dark matter halo rather than spherical one. In particular, Draco favours a strongly elongated dark matter halo, $Q \sim 0.2-0.3$, which are much more flattened than its observed stellar profile.
This outcome is naively in disagreement with the numerical simulations based on not only ULA dark matter but also the other possible dark matter scenarios even including concordance CDM models.

This result simply gives us a suggestion that dark matter haloes acquire a large angular momentum through their formation and evolution. 
However, in general, dark matter haloes formed in numerical simulations have a small spin parameter,\footnote{Spin parameter is defined as $\lambda=J/(\sqrt{2}MVR)$, where $M,R,V$ and $J$ are virial mass of the dark matter halo, radius, rotational velocity at $R$, and total angular momentum inside $R$.} $\log\lambda\sim-1.5$ in CDM models~\citep{1987ApJ...319..575B,2007MNRAS.376..215B,2013ApJ...767..146I} and $\log\lambda\sim-1.6$ in warm dark matter models~\citep{2016MNRAS.455..318B} on dwarf galaxy mass scales~($\sim10^{8-10}M_{\odot}$).
In addition, from observational facts~\citep[e.g.,][]{2008ApJ...688L..75W}, the dSphs have very weak stellar angular motions, thereby implying that their dark matter haloes would not have strong angular momenta.

On the other hand, \citet{2009ApJ...697..850W} performed the high-resolution ULA dark matter simulations and revealed that a quantum turbulence appears on the virialized boundary and could provide a triaxial density asphericity inside the halo due to the quantum anisotropic stress tensor.
They also calculated the degree of triaxiality of dark matter haloes and found that these ULA dark matter haloes have very weak triaxiality. In other words, the simulated ULA dark matter haloes would prefer to be rounder shapes.
In addition, \citet{2017MNRAS.471.4559M} have discussed the correlation between a turbulent peak power scale and a core size and showed that the turbulence actually becomes strong not inside a soliton core region but in outer regions of a dark matter halo. Thus, the triaxiality at the central core region would be weaker than an outer halo envelope.

\begin{figure}
	\begin{center}
		\includegraphics[scale=0.22]{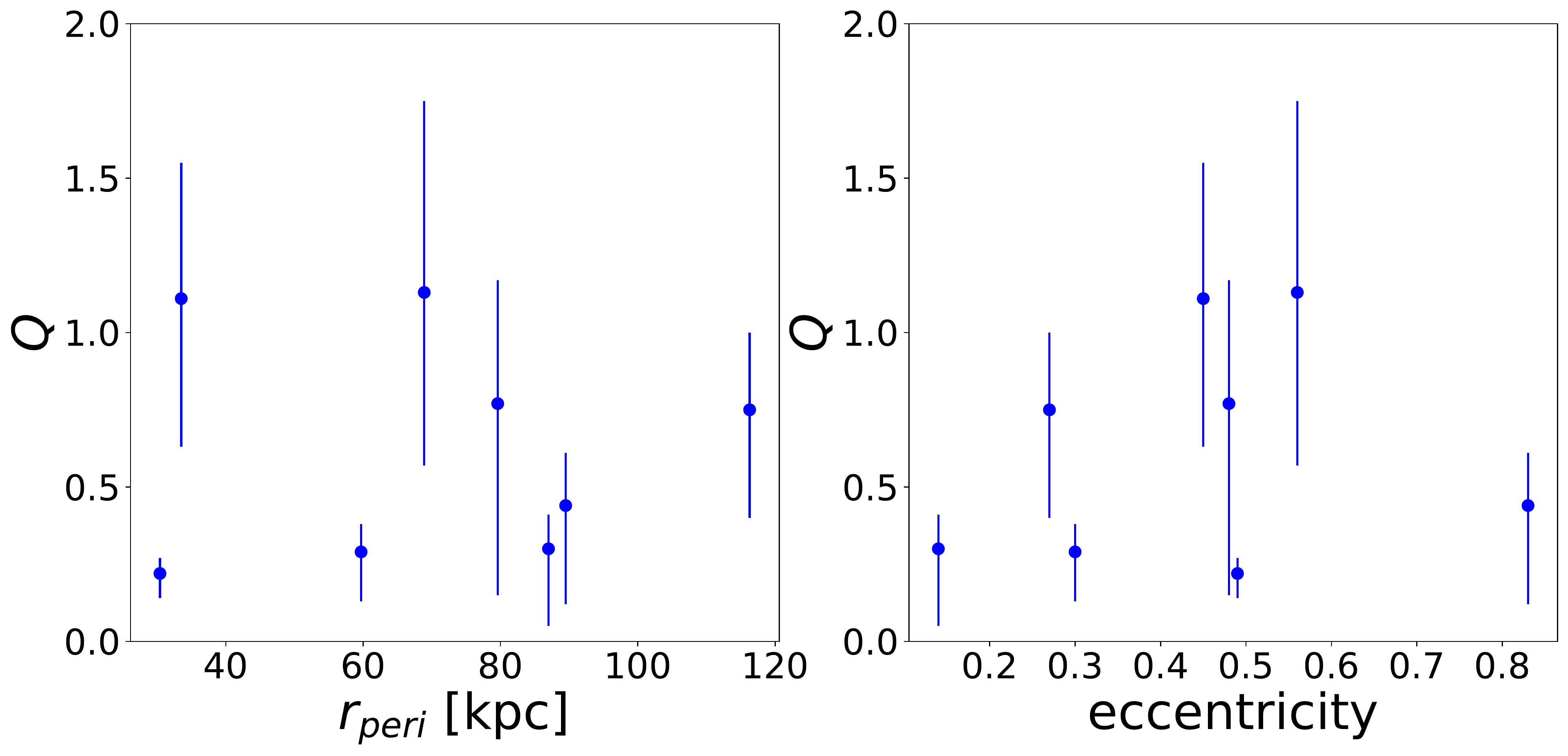}
	\end{center}
   \caption{The comparison between dark matter halo axial ratios~(for the case of a soliton core only model) and orbital parameters of the dSphs. The left panel shows $Q$~versus pericentre radius $r_{\rm peri}$, while the right one is $Q$ versus eccentricity of the orbit.}
    \label{orbits}
\end{figure}

Another possible mechanism is tidal effects from a host dark matter halo. Recent dark matter simulations present that subhaloes associated with a MW-sized host halo become elongated at pericentre, because these subhaloes are subject to the strong tidal force from a deep potential of their host dark matter halo~\citep[e.g.,][]{2007ApJ...671.1135K,2014MNRAS.439.2863V}.
However, the axial ratio of these simulated subhaloes is never smaller than $0.5$ even at the closest pericentric distance~($\sim50$~kpc) from the centre of their host halo~\citep[see Figure~5 in][]{2014MNRAS.439.2863V}.
In order to make a dark subhalo more flattened, we suggest a need for baryonic effects with respect to a host dark matter halo.
This is because adiabatic contraction of gas during disk formation makes a Milky Way-like dark matter halo potential deeper and steeper in its central part.
Then, subhaloes passing through such a deep potential of the host halo may have more flattened shapes by the stronger tidal distortions.
However, it is unclear how the baryonic effects indeed modify the shapes of dark subhaloes, especially the baryonic effects on a ULA dark matter halo have been largely unexplored yet.
Moreover, owing to recent results of {\it Gaia} satellite mission, the orbital parameters of the dSphs have become well known~\citep{2018A&A...616A..12G,2018A&A...619A.103F}.
Figure~\ref{orbits} denotes the comparison between dark matter halo axial ratios $Q$ and the orbital parameters~(pericentre and eccentricity) of the dSphs inferred by \citet{2018A&A...616A..12G}. From this figure, we find no remarkable relations within uncertainties of these parameters.

In light of the above suggestions, it could be challenging to explain the strongly elongated dark matter haloes of the dSphs in the framework of current dark matter models even with relying on baryon physics.
While further studies on the simulation fully considering baryonic effects are important to get an insight into the shapes of dark subhaloes, it would be also worth exploring the triaxial dynamical modeling of ULA haloes to fairly compare the shapes of dark matter haloes inferred from numerical simulations with those from observations.
We leave these issues in future works.

\section{Conclusions}
In this work, in order to obtain more reliable and realistic limits on ULA dark matter particle mass, we constructed non-spherical dynamical models of the soliton cored dark matter profile predicted by ULA dark matter simulations, and applied these models to the eight classical dSphs in the MW.
We found that the best-fitting cases for most of the dSphs yield not spherical but flattened haloes, even though there are some degeneracies between the axial ratio of dark matter haloes and the other free parameters.
This is because a fixed soliton core dark matter halo often requires a non-negligible change of a dark matter halo axial ratio $Q$ or a stellar anisotropy parameter $\beta_z$ as well as a soliton core radius $r_c$ to reproduce kinematic data.
In particular, Draco dSph prefers to be an extremely flattened shape of dark matter halo.
This is because a flattened dark matter halo shape~(i.e., $Q<1$) can roughly reproduce a flat or an upward line-of-sight velocity dispersion profile.
However, when a dark matter halo becomes more elongated such as $Q<0.5$, the features of $\sigma_{\rm los}$ profiles change drastically, and thus these dispersion profiles are quite different from the observed ones.
However, in combination with the effects of a tangentially-biased velocity anisotropy~($\beta_z<0$) and a large core radius of ULA dark matter halo~($r_c/b_{\ast}>1$), the shapes of dispersion profiles computed by a strongly elongated dark matter halo are in good agreement with the observed profiles especially for Draco.
Consequently, a flattened dark matter halo could be the most promising way to realize the observed line-of-sight velocity dispersion profiles.

We also made an attempt to place a constraint on particle mass of ULA, $m_{\psi}$, and found the combined $1\sigma$~$(2\sigma)$ confidence intervals of $\log_{10}([m_{\psi}/10^{-23}{\text{eV}}])=0.94^{+0.51 (+0.75)}_{-0.42 (-0.61)}$.
This combined constraint of $m_{\psi}$ is dominated by Fornax and Sculptor which have a large number of kinematic samples.
In comparison with the other Jeans analysis works, our resultant mass limit is less stringent than those by the previous spherical works because our mass models fully take into account the uncertainties of dark matter halo shape and the inclination angle.
Moreover, our constraint is in good agreement with the other independent constraints such as rotation curves of LSBs, CMB, the luminosity function of high-$z$ galaxy, and thickness of the MW stellar streams.

To justify the assumption that the member stars of the dSphs might reside within the central soliton core, we also developed the mass model where the soliton core connects to an NFW halo at a larger radius and applied this model to all dSphs.
Most of the dSphs~(Draco, Carina, Sextans, Leo~II and Sculptor) have larger transition radii~($r_\epsilon$) than the stellar half-light radii~($b_{\ast}$) and the radii that encloses 90~per~cent of the spectroscopic sample~($r_{90}$).
Moreover, the five parameters~$(Q,r_c,m_{\psi},\beta_z,i)$ for these galaxies are roughly consistent with those estimated by the soliton-only models, even though they seem to be somewhat affected by an outer NFW density profile.
However, it is found that for the other galaxies~(Ursa Minor, Leo~I and Fornax) their transition radii are much smaller than their $b_{\ast}$ and $r_{90}$.
From their fitting results we also found that there exist two presumable dark matter profiles: one is a soliton core only and another one is a small soliton core~$+$~NFW profile with large scale radius.
In conclusion, to obtain reliable and realistic limits on the dark matter halo predicted from ULA dark matter models, we emphasize that it might be necessary to take into account an external NFW dark matter profile encompassing a soliton core of dSphs.

Finally, we discovered that some of obtained axial ratios in dSphs are much smaller than theoretical predictions.
Although we discussed the possible mechanisms of a strongly elongated dark matter halo on sub-galactic scales, it could be challenging to explain such halo shapes in the framework of ULA dark matter models, so far.
Moreover, recent several studies of ULA dark matter through less-massive galaxies have argued that this dark matter model is inconsistent with the dark halo mass function of dSphs in the Milky Way~\citep{2019arXiv190611848S} and cannot reproduce the rotation curves of all SPARC galaxies~\citep{2018PhRvD..98h3027B,2019MNRAS.483..289R}.
However, there is still plenty of room to improve dynamical modelings for the dSphs. 
As an example, to compare dark matter halo shapes from observations with those from theoretical simulations adequately, further observational studies will be necessary to consider a triaxial shape of a system~\citep{2018MNRAS.476.2918K}.
Moreover, we should bear in mind that there are still large uncertainties on the inner slopes of dark matter profiles in the dSphs because of a small number of kinematic sample.
Therefore, the debate as to whether the central regions of dark matter haloes in dSphs are cored or cusped is still ongoing.
The future spectroscopic surveys such as the Prime Focus Spectrograph~\citep[][]{2016SPIE.9908E..1MT} attached to the Subaru Telescope~\citep{2014PASJ...66R...1T} will enable us to measure a large number of kinematic data for resolved stars in the dSphs and thus offer an opportunity to determine robustly the dark matter distributions in the dSphs. 

\section*{Acknowledgements}
We are grateful to the referee for her/his careful reading of our paper and thoughtful comments.
We would like to give special thanks to Masahiro Ibe, Masahiro Kawasaki, Shigeki Matsumoto, Evan Kirby, Rosemary Wyse, and Masashi Chiba for useful discussions.
This work was supported in part by the MEXT Grant-in-Aid for Scientific Research on Innovative Areas, No.~18H04359 and No.~18J00277~(for K.H.).
Numerical computations were carried out on Cray XC50 at Center for Computational Astrophysics, National Astronomical Observatory of Japan.

\appendix
\section{Effects of the parameters on line-of-sight velocity dispersion profile}
In this appendix, we demonstrate the impacts of a non-spherical shape of dark halo, $Q$, a velocity anisotropy, $\beta_z$, and a ratio of $r_c/b_{\ast}$ on line-of-sight velocity dispersion profiles.

\begin{figure*}
	\begin{center}
		\includegraphics[scale=0.67]{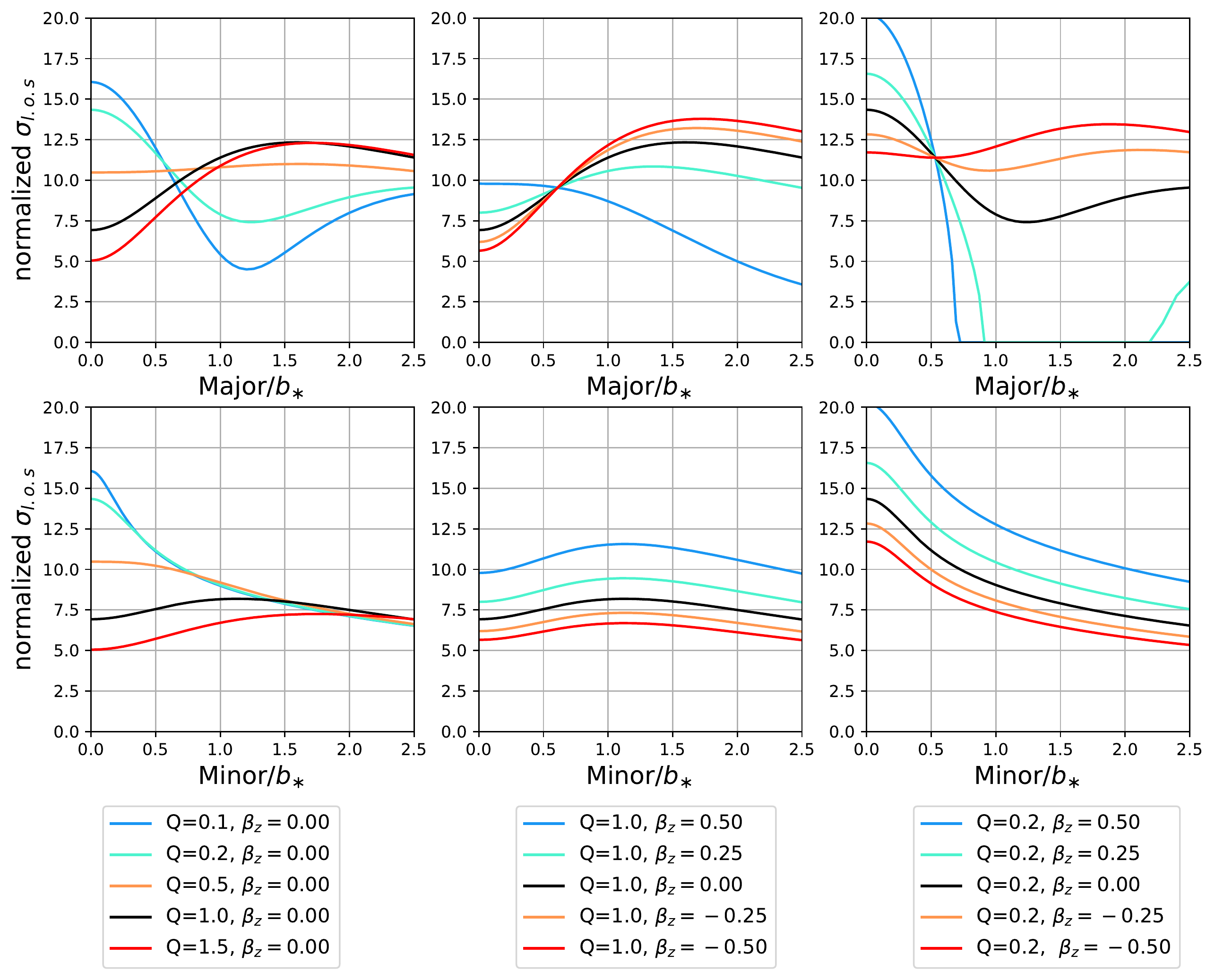}
	\end{center}
   \caption{Upper panels show the normalized line-of-sight velocity dispersions, $\sigma_{\rm l.o.s}/(Gb^2_{\ast}\rho_{c,r_{c}/b_{\ast}=3})^{1/2}$, along the major axis for soliton core only case, whereas the lower panels show these velocity dispersions along the minor axis for the same case. 
   {\it Left column:} line-of-sight velocity dispersion profiles with changing $Q$ under $\beta_z=0$,
   {\it center column:} those with changing $\beta_z$ under $Q=0$, and
   {\it right column:} those with changing $\beta_z$ under $Q=0.2$.
   For all of these cases we suppose that the oblate stellar distribution~($q=0.7$), the edge-on galaxy~($i=90^{\circ}$), $m_{\psi}=10^{-22}$~eV  and the ratio of $r_{c}/b_{\ast}=1$ for the sake of demonstration.}
    \label{demolos}
\end{figure*}

\begin{figure*}
	\begin{center}
		\includegraphics[scale=0.67]{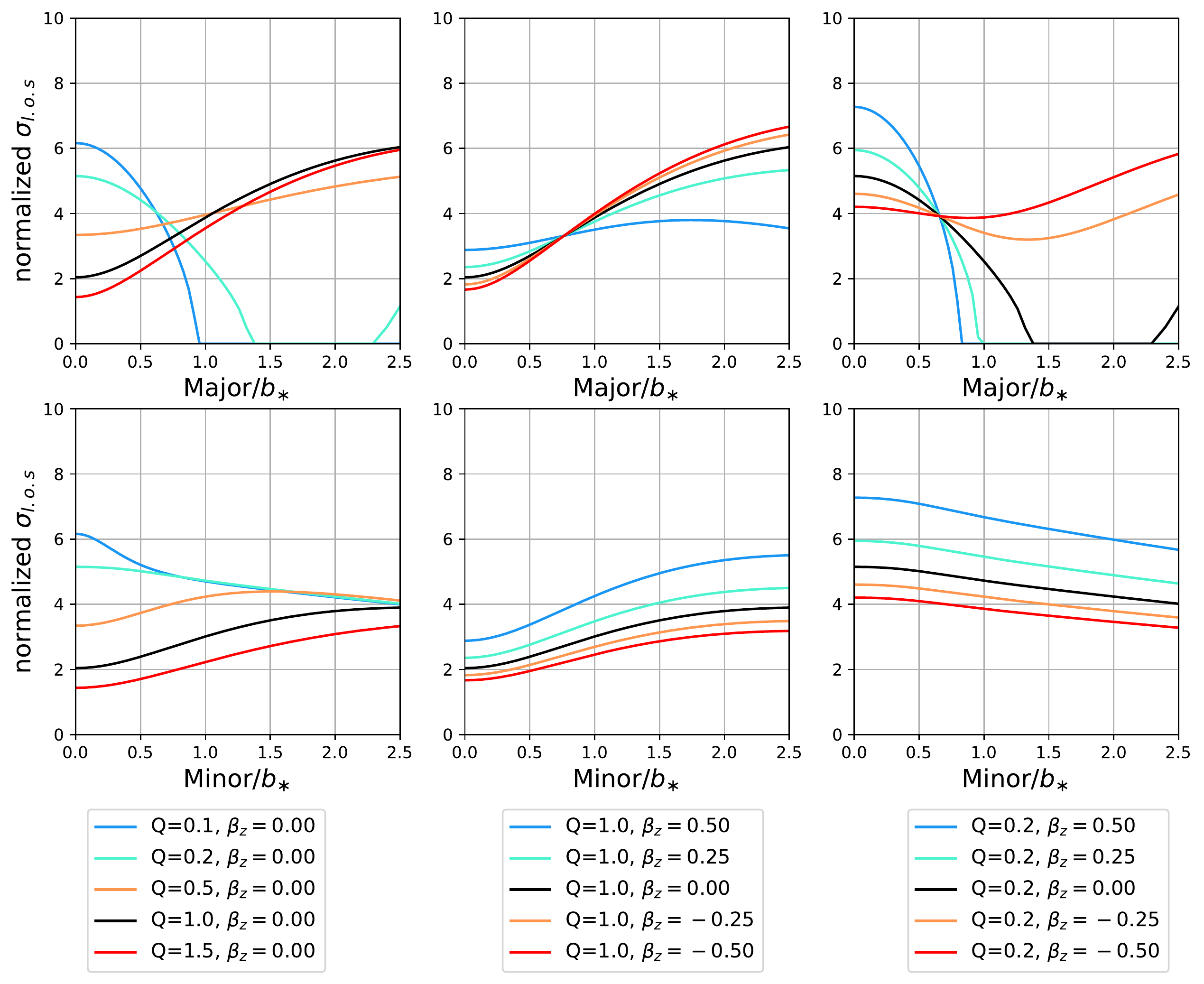}
	\end{center}
   \caption{Same as figre~\ref{demolos}, but for $r_{c}/b_{\ast}=2$}
    \label{demolos2}
\end{figure*}

\begin{figure*}
	\begin{center}
		\includegraphics[scale=0.67]{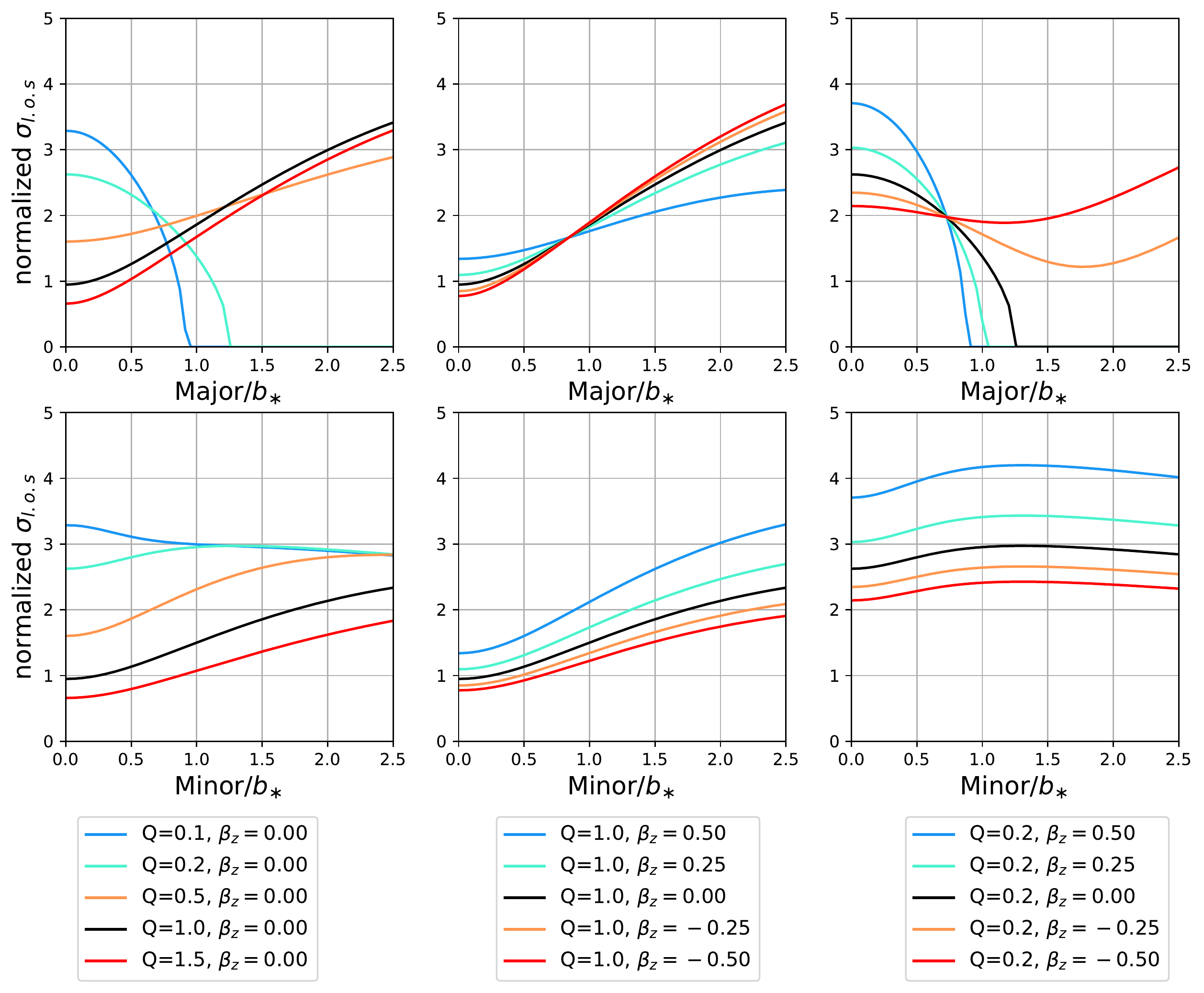}
	\end{center}
   \caption{Same as figre~\ref{demolos}, but for $r_{c}/b_{\ast}=3$}
    \label{demolos3}
\end{figure*}

Figure~\ref{demolos}-\ref{demolos3} depict the normalized line-of-sight velocity dispersions along the major and minor axes, respectively.
The ratio of $r_c/b_{\ast}$ differs in each figure. 
For comparison, all of the velocity dispersions are normalized by $(Gb^2_{\ast}\rho_{c,r_{c}/b_{\ast}=3})^{1/2}$, where $\rho_{c,r_{c}/b_{\ast}=3}$ is the central density in the case for $r_{c}/b_{\ast}=3$. 
We set the fixed axial ratio of a stellar distribution in this test calculation, $q=0.7$, which is a typical value of the Galactic dSphs.
We also fix the inclination angle at $i=90^{\circ}$ and ULA dark matter mass at $m_{\psi}=10^{-22}$~eV for the sake of demonstration.
In the left panels we change the value of $Q$ with no velocity anisotropy $\beta_z=0$, while we change $\beta_z$ under the spherical dark matter halo $Q=1$ in the middle panels.
Moreover, the right panels show that $\beta_z$ are changed under the elongated shape of dark matter halo $Q=0.2$.

First, as is shown in the upper panels of Figure~\ref{demolos} denoting the line-of-sight velocity dispersion profiles along the major axis, we can see that given that a galaxy has an oblate stellar distribution (i.e., stellar axial ratio is less than unity,~$q<1$), their $\sigma_{\rm los}$~profiles have wavy features from the central to outer parts of the galaxy~\citep[see also Figure 12 in][]{2015ApJ...810...22H}.
In the top left panel of Figure~\ref{demolos}, the effect of decreasing $Q$ from unity to $0.5$ weakens the wavy feature of the $\sigma_{\rm los}$~profile cased by a non-spherical $Q$ along the major axis, as already discussed in~\citet{2012ApJ...755..145H}.
However, when a dark matter halo becomes more elongated such as $Q<0.5$, the wavy feature is transformed from trough- and crest-like feature into crest- and trough-like one.
In the top middle panel of Figure~\ref{demolos}, the effects of radially-biased velocity anisotropy~(i.e. $-\log(1-\beta_z)>0$) are resemblant to those of decreasing $Q$. 
In other words, a radially-biased velocity anisotropy is capable to increase (decrease) a $\sigma_{\rm los}$ profile in inner (outer) parts of a galaxy.
By contrast, tangentially-biased ones can enhance the wavy feature. 
Finally, the top right panel of Figure~\ref{demolos} shows the $\sigma_{\rm los}$~profiles along major axis in the case for very flattened dark matter halo ($Q=0.2$) with changing $\beta_z$.
Focusing on the effect of radially-biased velocity anisotropy, these $\sigma_{\rm los}$ profiles have {\it unphysical} region where $\sigma_{\rm los}$ becomes zero or negative values.
This is because this effect can help decrease $\sigma_{\rm los}$ profiles in outer parts of a galaxy.

Second, as is shown in the lower panels of Figure~\ref{demolos}, we can see that these $\sigma_{\rm los}$ profiles along the minor axis have a couple of features different from those along the major axis.
In the lower left panel, the effect of decreasing $Q$ increases $\sigma_{\rm los}$ at only inner parts. This is because a smaller $Q$ yields stronger gravitational force in the $z$-direction, thereby increasing $\overline{v^2_z}$ in inner parts.
On the other hand, in the lower middle panel, the effect of changing $\beta_z$ is monotonous, that is, this mainly affects the amplitude of the $\sigma_{\rm los}$ profiles and only changes it overall shape weakly. 
This is caused by $\sigma_{\rm los}$ along the minor axis, which is not contributed by $\overline{v^2_{\phi}}$.
From the aforementioned explanations, the trend of $\sigma_{\rm los}$ profiles in the lower right panel can be straightforwardly understood.

Third, comparing between Figure~\ref{demolos}, \ref{demolos2}, and \ref{demolos3}, the effect of increasing $r_c/b_{\ast}$ ratio decreases the amplitude of $\sigma_{\rm los}$ profiles along both axes.
This can be easily understood because the central density $\rho_c$ is proportional to $r_c^{-4}$ from Equation~(\ref{solitoncore}).
Furthermore, in the upper panels in each Figure, we can see that the wavy features of $\sigma_{\rm los}$ profiles are extended slightly toward the outside of a galaxy with the increasing $r_c/b_{\ast}$ ratios. 
Therefore, the effect of increasing $r_c/b_{\ast}$ weakens the wavy feature of $\sigma_{\rm los}$ profiles along the major axis.

In summary, we inspect the effects of $Q$, $\beta_z$, and $r_c/b_{\ast}$ on line-of-sight velocity dispersion profiles along the major and minor axes.
From this analysis, we show that a flattened stellar system has a wavy feature of its $\sigma_{\rm los}$ profile from the central to outer parts of the system.
Also, it is found that changing the values of $Q$, $\beta_z$, and $r_c/b_{\ast}$, this wavy feature can be both weakened and strengthened.

\section{The best-fit ULA $+$ NFW dark matter halo profiles}
\begin{figure*}
	\includegraphics[scale=0.34]{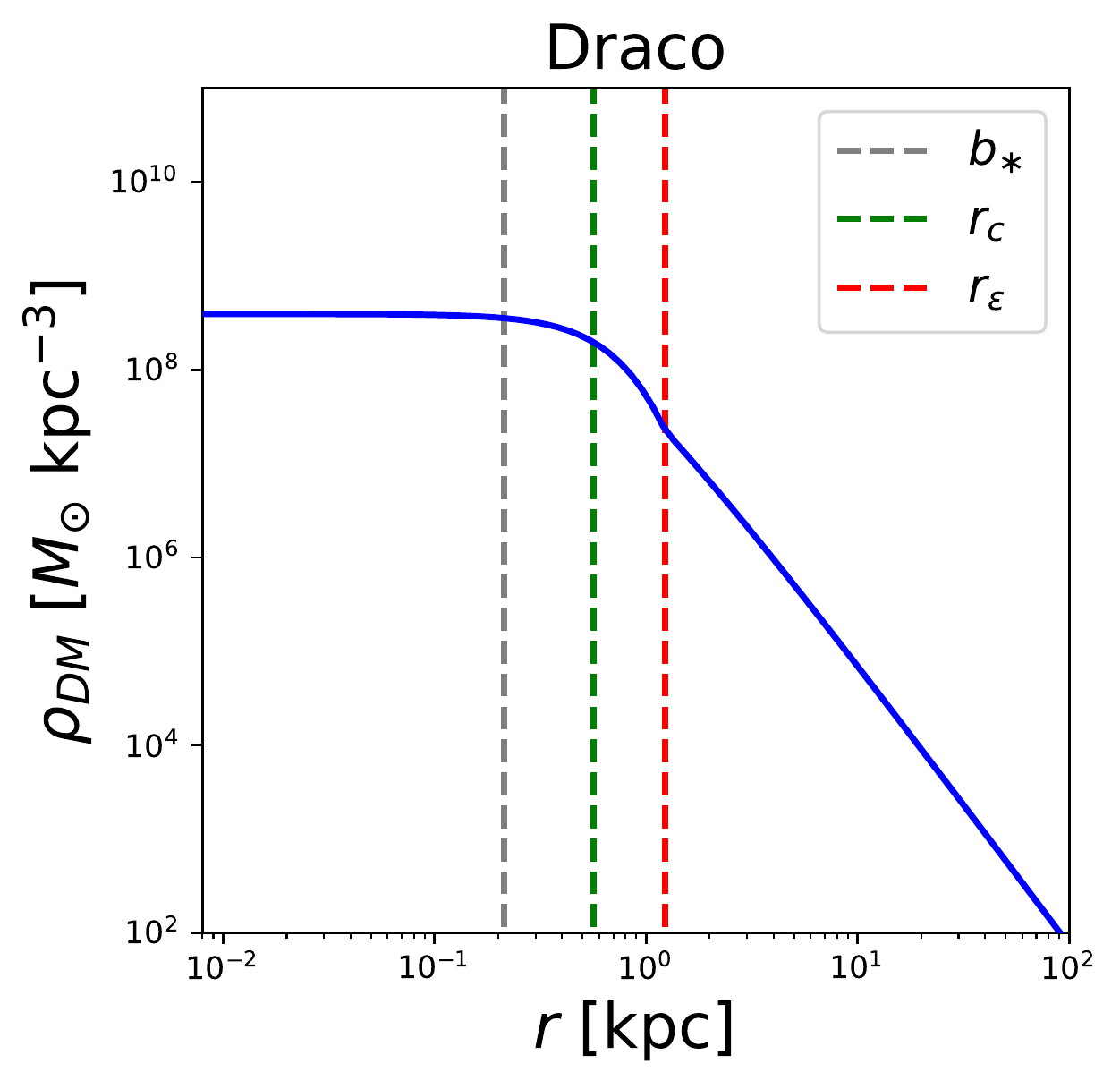}
	\includegraphics[scale=0.34]{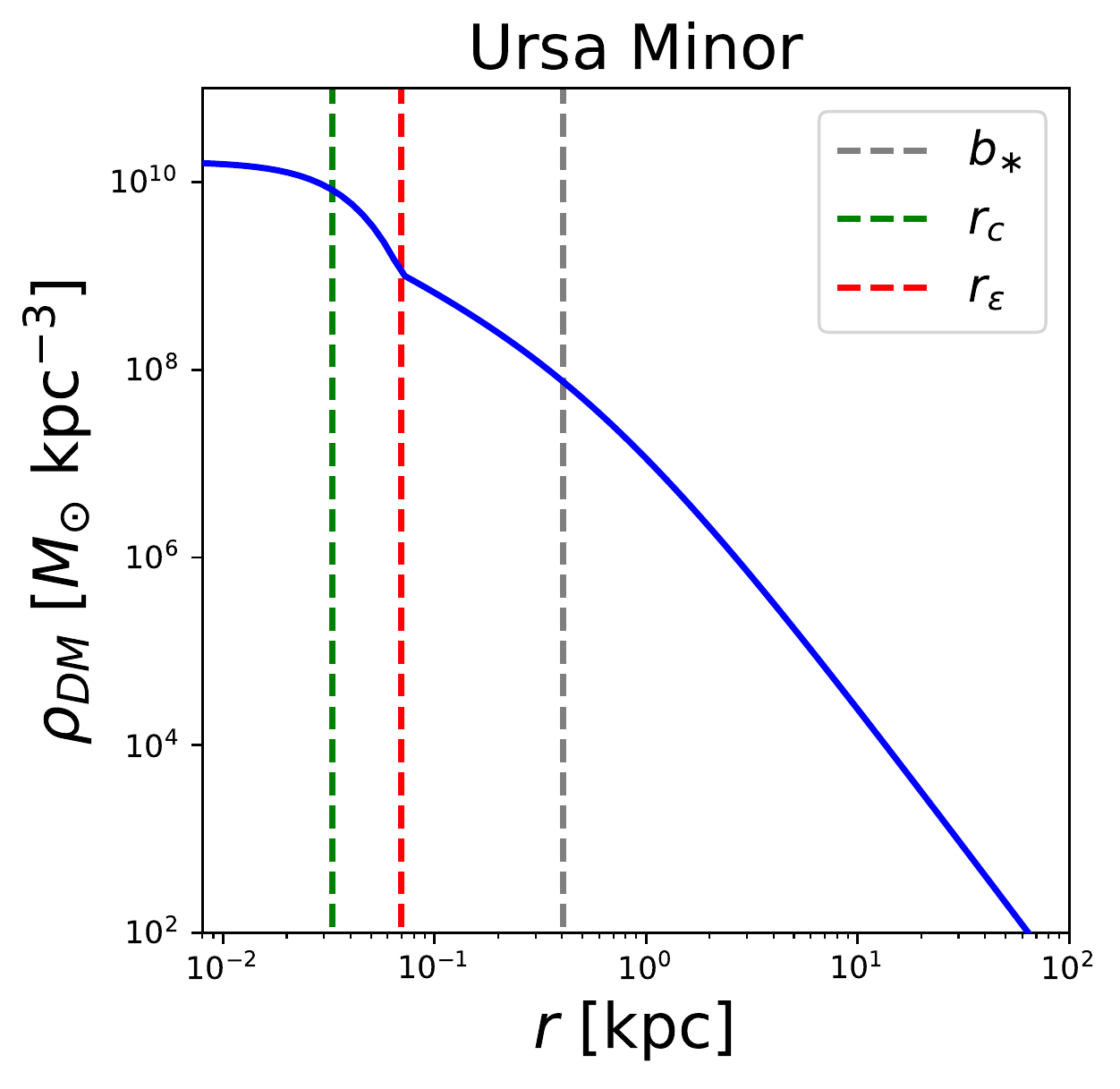}
	\includegraphics[scale=0.34]{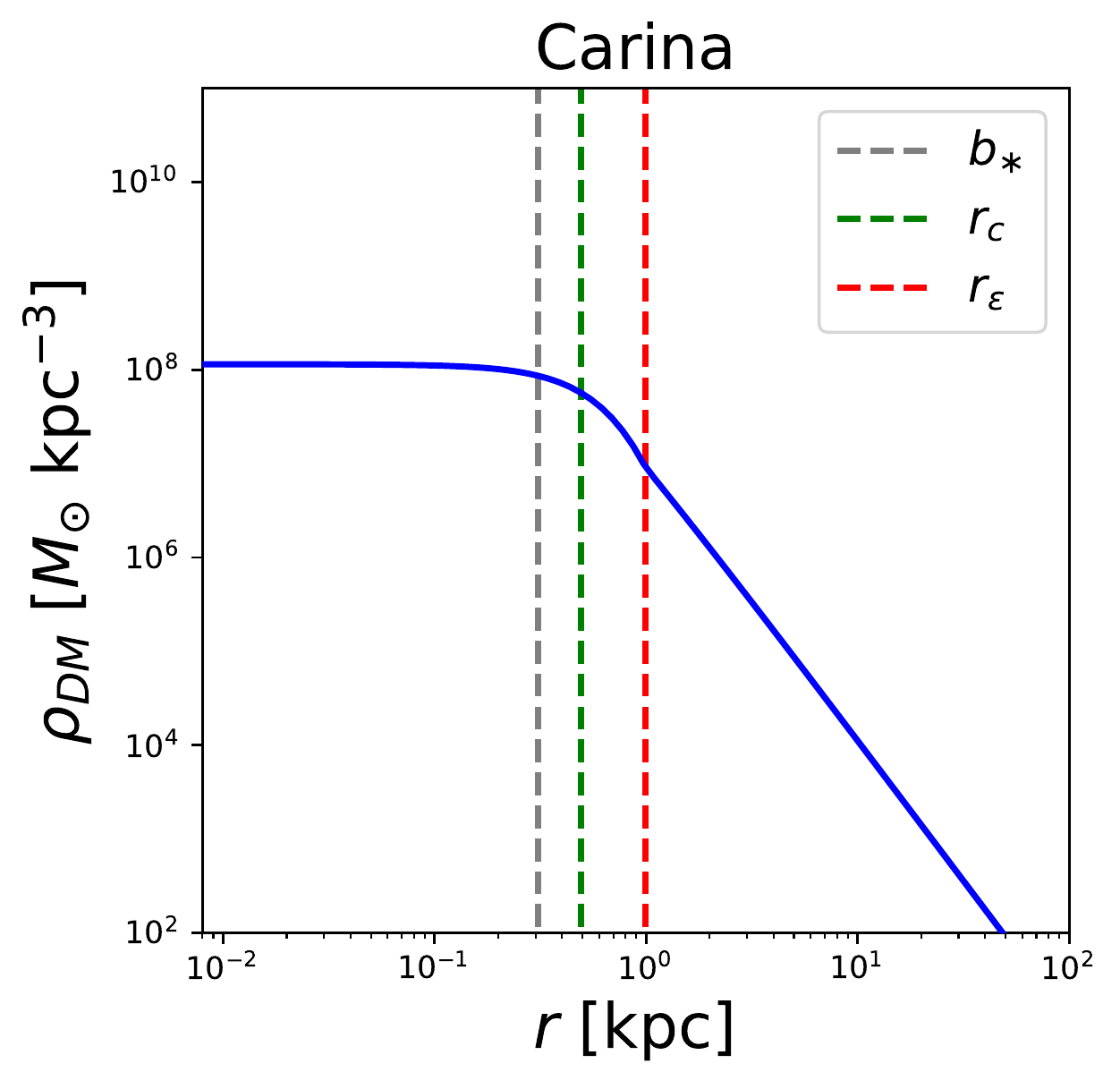}
	\includegraphics[scale=0.34]{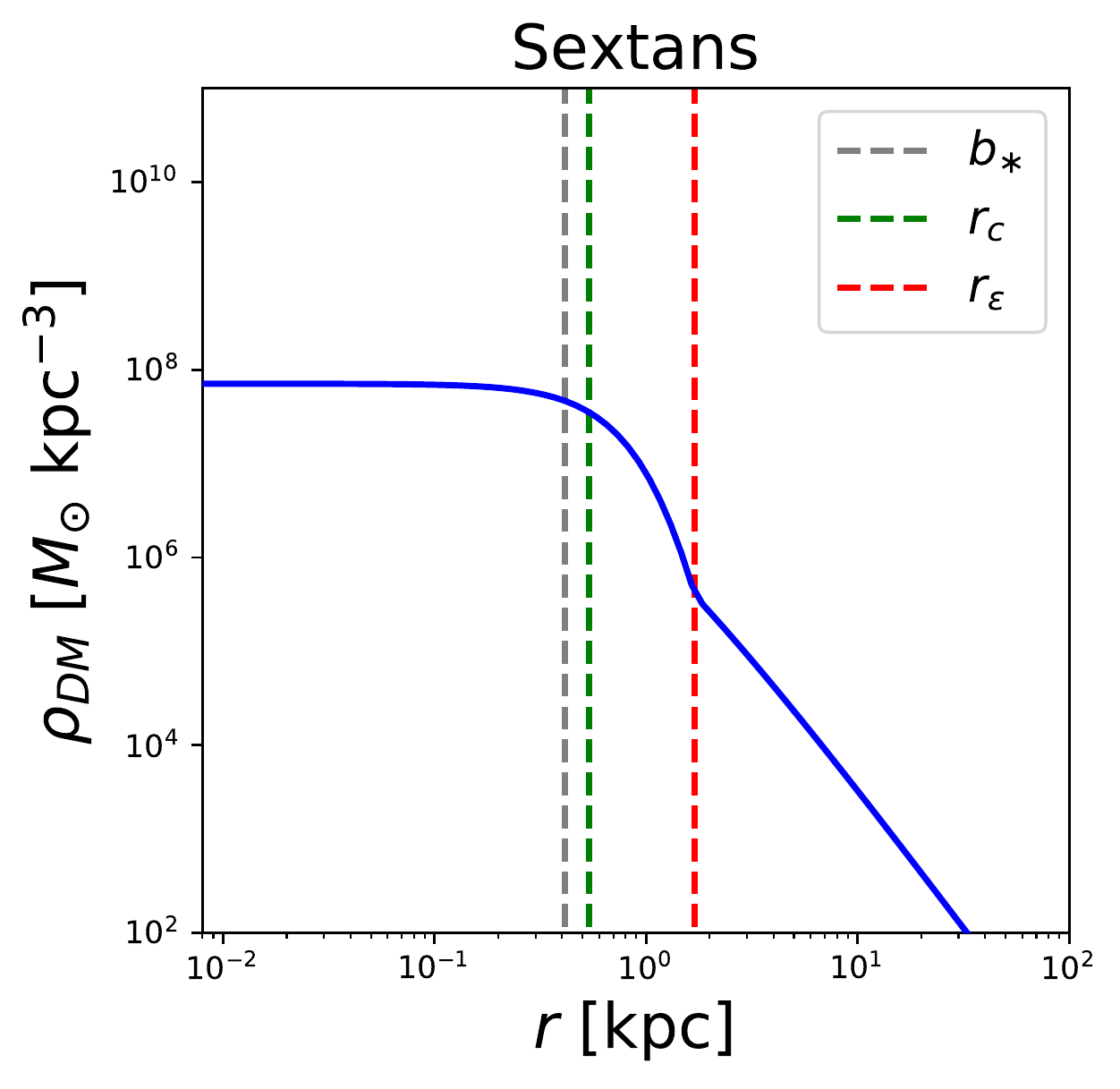}
	\includegraphics[scale=0.34]{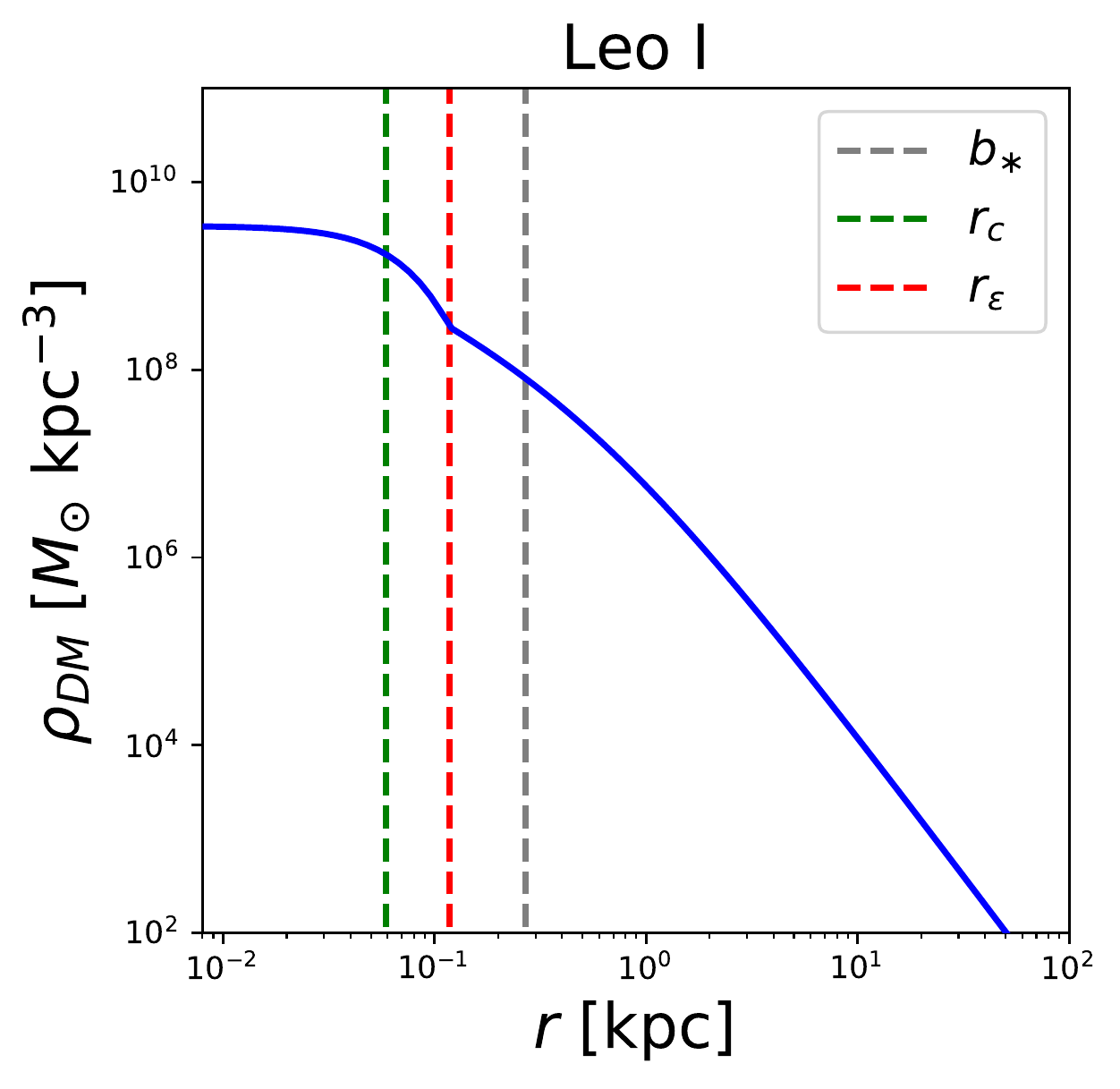}
	\includegraphics[scale=0.34]{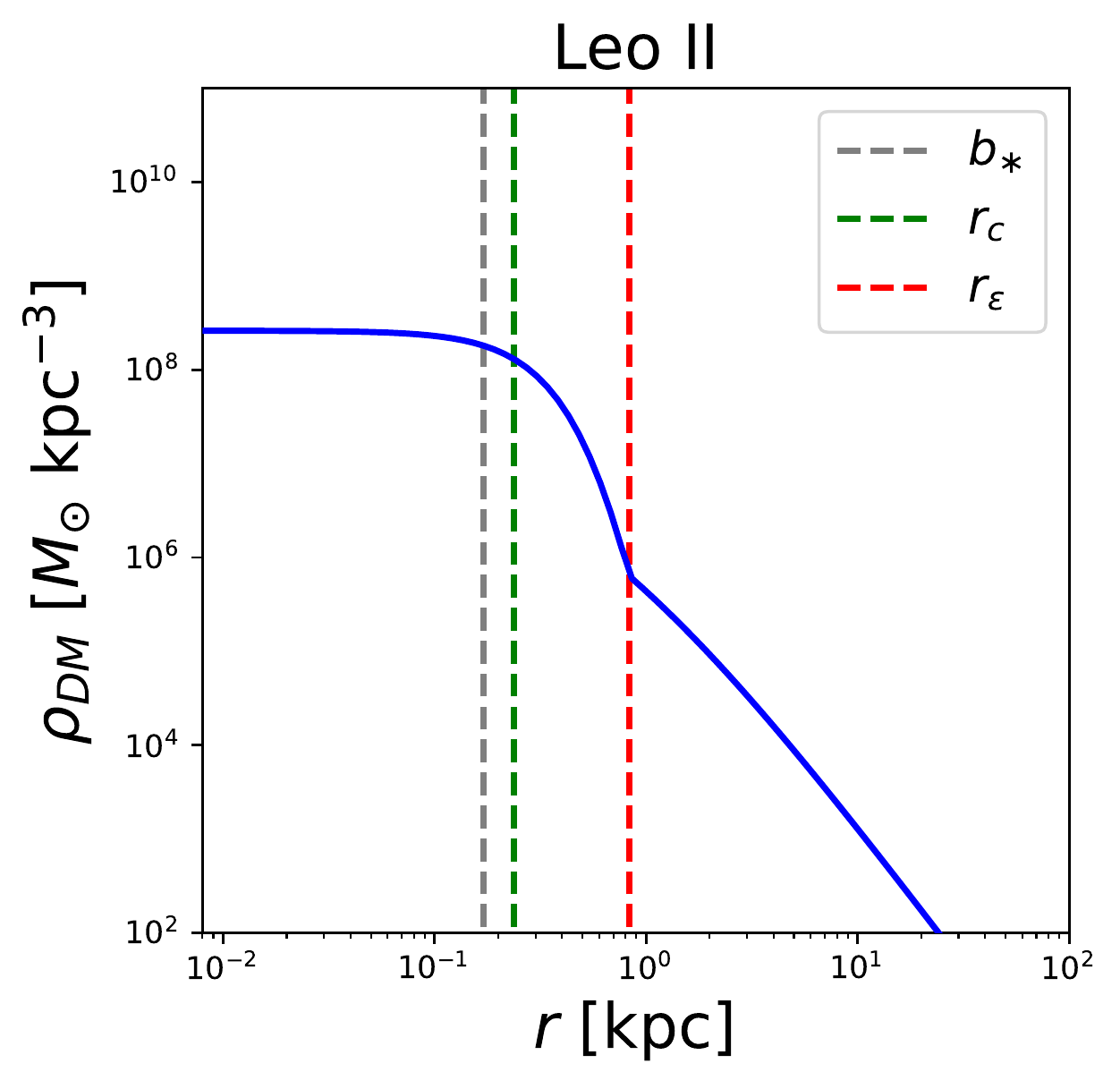}
	\includegraphics[scale=0.34]{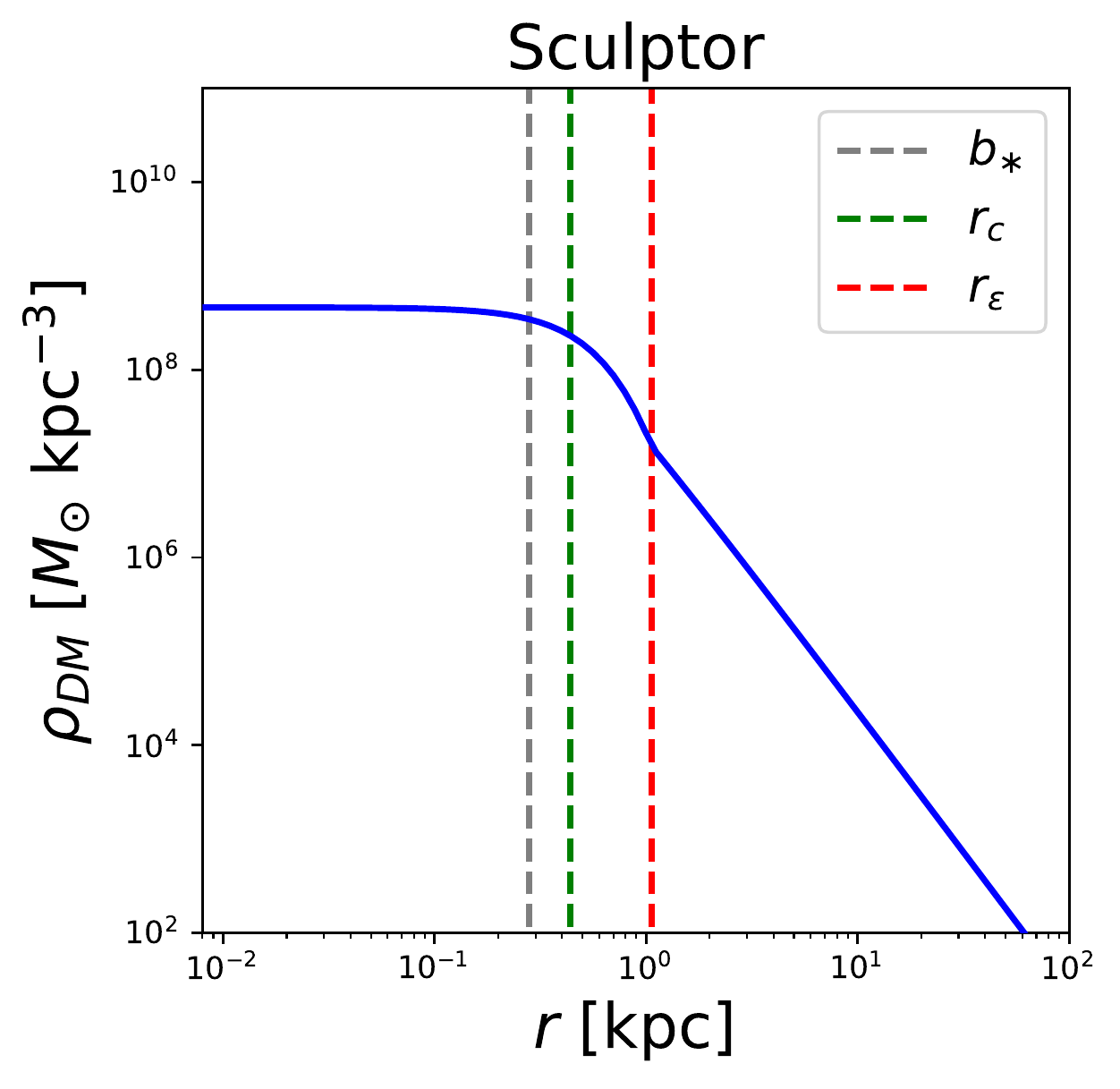}
	\includegraphics[scale=0.34]{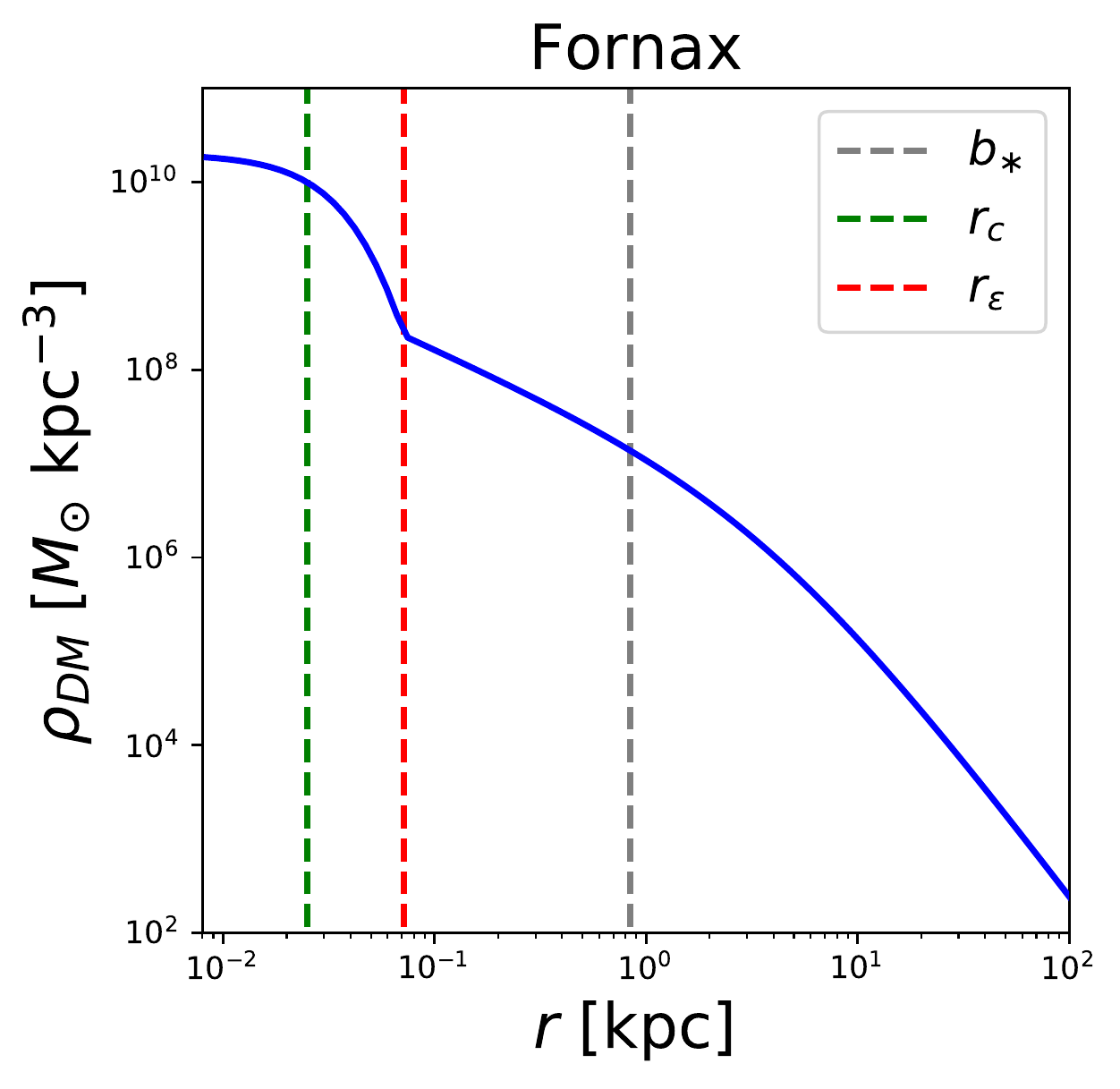}
    \caption{ULA and NFW dark matter density profiles of the dSphs computed from the best-fit parameters~(Table~\ref{table3}). The gray, green and red dashed lines denote a half-light radius, a soliton core radius and a transition radius from soliton to NFW density profiles, respectively.}
    \label{DMpro}
\end{figure*}

Figure~\ref{DMpro} shows ULA and NFW dark matter density profiles of the dSphs computed from the best-fit parameters~(Table~\ref{table3}).
From this figure, the transition radii, $r_{\varepsilon}$, for most of the dSphs are much larger than their half-light radii, $b_{\ast}$.
By contrast, Ursa~Minor, Leo~I and Fornax dSphs have smaller $r_{\varepsilon}$ than  $b_{\ast}$.
These galaxies could prefer to have a small soliton core or NFW cusped profile with a large scale radius.

\bibliographystyle{mnras}
\input{manuscript.bbl}

\bsp	
\label{lastpage}
\end{document}